\documentclass[12pt]{article}
\pdfoutput=1
\usepackage[nosort]{cite}
\usepackage[normalem]{ulem}
\usepackage{lipsum,calligra,pbsi}

\usepackage[dvipsnames]{xcolor}
\usepackage{epsfig}
\usepackage{amsfonts}
\usepackage{amscd}
\usepackage{latexsym}
\usepackage{amsmath, amssymb}
\usepackage{verbatim}
\usepackage{color}
\usepackage{fancyhdr}
\usepackage{hyperref}
\usepackage{tikz}
\usepackage{slashed}
\usepackage{multirow}

\usetikzlibrary{calc}
\usetikzlibrary{topaths}
\usetikzlibrary{decorations}
\usetikzlibrary{decorations.pathmorphing}
\usetikzlibrary{arrows,decorations.markings,cd}
\usetikzlibrary{calc,arrows,cd,decorations.markings,snakes}
\tikzset{
->-/.style args={#1rotate#2}{decoration={markings, mark=at position #1 with {\arrow[scale=1.5,rotate = #2 ]{stealth}}}, postaction={decorate}}
}
\usetikzlibrary{shapes.geometric}
\usetikzlibrary{knots}
\usepackage{tikz-3dplot}

\usepackage{float}
\usepackage{draft}
\usepackage{graphicx, subfig}
\usepackage{mciteplus}
\usepackage{skak}
\usepackage{bbm}

\usepackage{cleveref}
\usepackage{algorithm}
\usepackage[letterpaper,left=1.1in,top=1in,right=1.1in,bottom=1.3in]{geometry}
\usepackage{makecell}
\usepackage{enumerate}
\usepackage{tablefootnote}
\usepackage{longtable}
\usepackage{mathrsfs}

\numberwithin{equation}{section}

\usepackage[us,hhmmss]{datetime}

\usepackage{pifont}

\usepackage[scr=euler]{mathalfa}

\usepackage{subcaption}
\usepackage[english]{babel}
\usepackage{soul}
\usepackage{upgreek}
\usepackage[scr]{rsfso}

\usepackage{tikz}

\usepackage{longtable}

\DeclareMathAlphabet{\zapfcal}{OT1}{pzc}{m}{it}

\DeclareFontFamily{OT1}{pzc}{}
\DeclareFontShape{OT1}{pzc}{m}{it}{<-> s * [1.10] pzcmi7t}{}
\DeclareMathAlphabet{\mathpzc}{OT1}{pzc}{m}{it}

\numberwithin{equation}{section}

\colorlet{mylinkcolor}{NavyBlue}
\colorlet{mycitecolor}{Aquamarine}
\colorlet{myurlcolor}{Aquamarine}
\newcommand\myshade{90}

\hypersetup{
  linkcolor  = mylinkcolor!\myshade!black,
  citecolor  = mycitecolor!\myshade!black,
  urlcolor   = myurlcolor!\myshade!black,
  colorlinks = true,
}

\def\U{\mathrm{U}(1)}

\def\cA{\mathcal{A}}

\def\bZ{\mathbb{Z}}
\def\bR{\mathbb{R}}
\def\cT{\mathcal{T}}

\def\U{\mathrm{U}(1)}

\begin{document}
\begin{titlepage}

\title{Proliferation transitions from a topological phase in $2+1$ dimensions}

\author{Meng Cheng$^{1,2}$ and Nathan Seiberg$^2$}

 \address{${}^1$ Department of Physics, Yale University, New Haven, Connecticut 06511, USA}
 \address{${}^2$ School of Natural Sciences, Institute for Advanced Study, Princeton, NJ}
 
\abstract{
\noindent   We consider phase transitions out of a general topological phase in $2+1$ dimensions.  We assume that the transition is triggered by a single Abelian anyon, which becomes light near the transition and whose worldlines proliferate after the transition.  (This proliferation is often referred to as ``condensation.'') We describe the transition using a continuum field theory obtained by coupling the corresponding topological quantum field theory (TQFT) to a single complex scalar field associated with this anyon.  With these assumptions, we find the most general relativistic field theory for such a transition.  Even though for a given TQFT and a choice of anyon, there are infinitely many such field theories, the transition theory depends on only a single additional integer parameter.  We analyze all these theories, their global symmetries, and their phases. In generic cases, the theory after the transition can be related to the original one via an Abelian hierarchy construction. In special cases, the theory after the transition is gapless, and with a particular deformation, it is related to the original TQFT by gauging an anomaly-free one-form global symmetry.  We also explore the enrichment of this setup by a global $\U$ symmetry.  In some cases, enriching the original TQFT is incompatible with the full transition theory.  Finally, we demonstrate our construction with many specific examples.} 
\end{titlepage}
\tableofcontents

\section{Introduction}\label{Introduction}

\begin{figure}
\centering
\includegraphics[width=15cm]{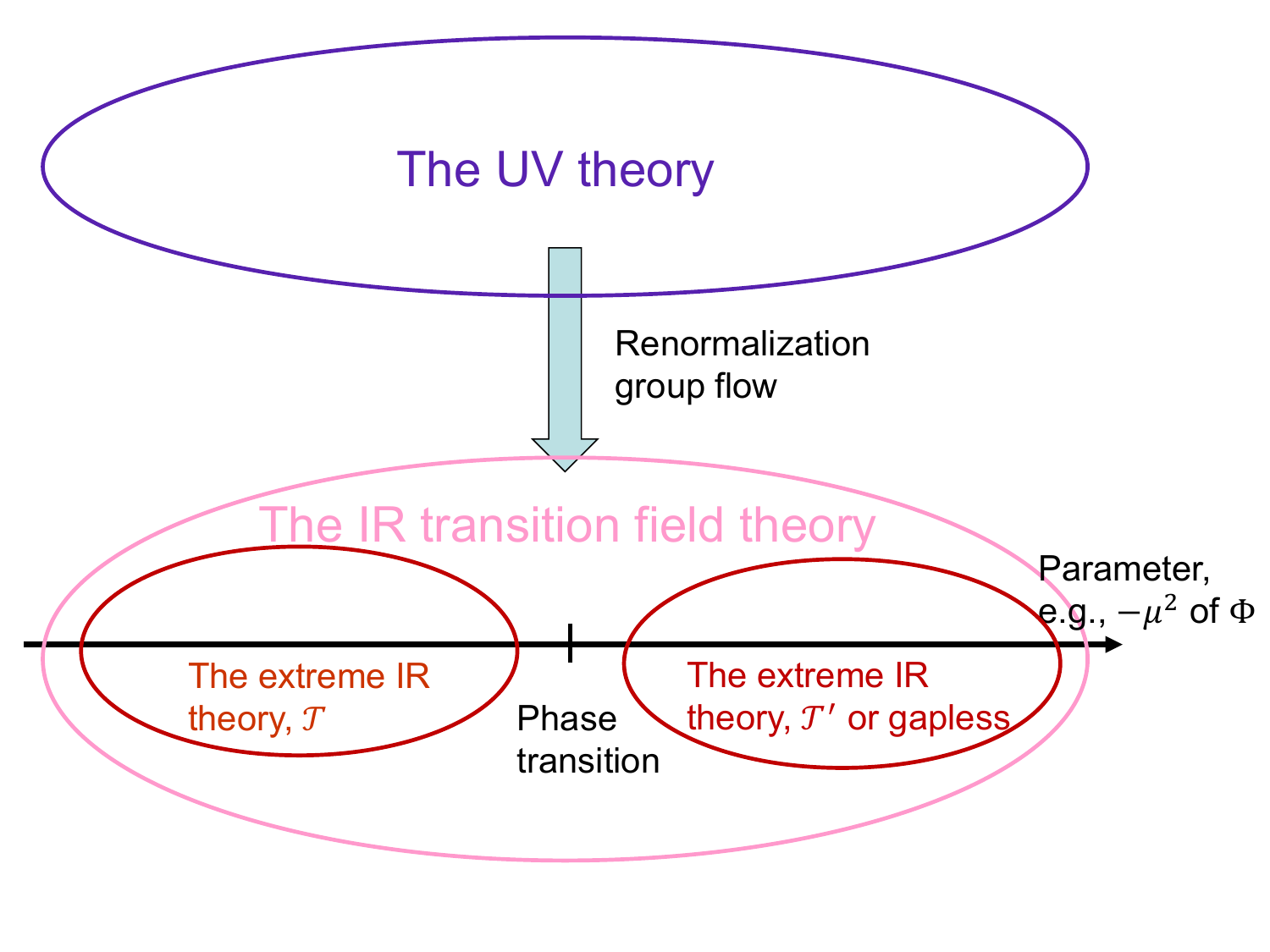}
\caption{Theories at three different scales.  At very short distances, denoted in violet, we have a UV theory.  It can be either a lattice or a continuum theory. In Section \ref{enrichmentsec}, we will examine the consequences of a global $\U_A$ symmetry acting at that scale.  We follow the renormalization group flow to the IR.  In the extreme IR, denoted in red, we consider two phases.  One of them is described by a TQFT $\cal T$.  The other phase is either a TQFT $\cal T'$ or a gapless phase.  The main focus of this note is the transition continuum field theory that operates at intermediate energies and is denoted in pink.  It captures the two extreme IR phases and the phase transition between them.  We assume that in addition to various topological modes, this theory includes only a single complex scalar field $\Phi$. This scalar field creates an anyon $a$ in the phase with $\cal T$.  The transition happens when we dial the mass square $\mu^2$ of $\Phi$ from positive to negative.  As we do that, the anyon $a$ becomes light, and its worldlines proliferate in the other phase.  In terms of $\Phi$, this is a Higgs transition.}
\label{threescales}
\end{figure}

The goal of this paper is to analyze phase transitions out of a given $2+1$-dimensional gapped phase described by a topological quantum field theory (TQFT) $\cal T$.  This problem has been studied by many authors.  The novelty here will be a more systematic search for a transition field theory and the detailed analysis of its properties.

One way this problem can arise is by starting with a short-distance UV theory, whether on the lattice or in the continuum.  If the system is gapped, its low-energy dynamics is described by a TQFT $\cal T$.  As we vary the parameters in the UV theory, a phase transition can occur to another phase, which can be another gapped phase with a TQFT $\cal T'$ or a gapless phase.

Clearly, this transition can be described in the full UV theory.  A more efficient description focuses on intermediate energies, far below the UV scale and far above various light modes.  Such a description can capture the two phases and the transition between them, but it is independent of most of the details of the UV theory.  We will refer to this theory as the transition field theory.  This situation involving the UV theory, the intermediate-energy transition field theory, and the extreme IR theories is depicted in Figure \ref{threescales}.

A natural setting for such a transition theory is the following.  The TQFT $\cal T$ can be described by a Chern-Simons theory \cite{Witten:1988hf}.  (Importantly, our analysis below does not rely on having such a Lagrangian description of  $\cal T$.  Instead, it can be viewed as an abstract theory.) Then, as we approach the transition, we should add non-topological modes.  These include some anyons that become light near the transition.  This leads us to study Chern-Simons-matter theories.   Chern-Simons-matter theories have been discussed by many people from different perspectives.  These include various supersymmetric and non-supersymmetric gauge theories, which exhibit rich dualities.  Closer to our discussion, there are also many papers that use Chern-Simons-matter theories to describe phase transitions of interest in condensed matter physics. See e.g., the early papers  \cite{kivelson1992global,WenWu1993,chen1993mott} and the more recent ones \cite{Barkeshli:2012rja, LeeQED3,Goldman:2019wvz, ZouQCD3,Goldman:2020khm, MaQCD3, JensenRaz, song2024phase,shi_analytic_2024,Ji:2026yfj,Dumitrescu:2026vre}.

In this note, we will limit ourselves to what is arguably the simplest set of such Chern-Simons matter theories. We add to the TQFT $\cal T$ a single complex scalar field $\Phi$.

In a TQFT, all the anyons are infinitely heavy, external probes, and they are described by line defects,\footnote{We will refer to them interchangeably as line operators, line defects, or simply lines.} which represent their worldlines. In a Chern-Simons description of the TQFT, based on a gauge group $G$, some of these lines are Wilson lines in some representations of $G$. Others can be defined as those that induce a particular holonomy of $G$ around them \cite{Witten:1988hf,Elitzur:1989nr,Moore:1989yh,Gukov:2008sn}.

We would like to focus on a single Abelian anyon $a$ and make it dynamical, so that it can trigger a transition.  Intuitively, this anyon proliferates on the other side of the transition.  More precisely, since the anyon is not gauge invariant, the objects that proliferate are the lines attached to it, rather than the anyons.

Some of these transitions map $\cal T$ to $\cal T'$, which is an Abelian hierarchy construction out of $\cal T$ \cite{HaldaneHierarchy, HalperinHierarchy}.  This is often described as ``condensation.''  Other transitions lead to $\cal T'$, which is related to $\cal T$ via gauging a one-form symmetry \cite{Moore:1989yh}. (See Appendix \ref{sec:gauging-one-form} for a review of gauging a one-form symmetry.) Following  \cite{Bais:2008ni}, these are often described as ``anyon condensation.''  We will refrain from using this terminology.\footnote{
The standard notion of ``condensation,'' as in the case of vapor turning into liquid, is the process of forming a state with dense particles.  The particle number is not spontaneously broken.  Often, as in the case of vapor turning into liquid, this is not an invariant characterization of the phase.  Another notion of condensation involves the spontaneous breaking of a global symmetry, e.g., particle number in a superfluid.  All the transitions discussed in this paper can be viewed as condensation transitions in the first sense.  In some of them, a global $\U$ symmetry is spontaneously broken.  But typically, this is not the case.  To avoid confusion, we will refer to all the transitions as proliferation transitions.\label{nocondensation}}

In a second-quantized picture, we make the anyon $a$ dynamical by introducing a field $\Phi$ that creates it.  Clearly, this field should not correspond to a well-defined local operator.\footnote{Here, we refer to operators that act at a point as local operators.  They can also be called point operators.  They are to be distinguished from line operators or surface operators.} Instead, it is the end of a line operator/defect -- the worldline of the anyon.    In a Chern-Simons description of the TQFT, where the line is a Wilson line of $G$ in some representation, the field $\Phi$ transforms as this representation of $G$.

This intuitive picture points to some ambiguities:
\begin{itemize}
\item The same TQFT can be described by different dual Chern-Simons theories based on different gauge groups $G$.  This could affect the description with $\Phi$.
\item Even for a given choice of gauge group $G$, the same anyon $a$ can correspond to different representations of $G$.  They differ by a transparent line in the TQFT.  
\item  As mentioned above, depending on the gauge group $G$ and the anyon $a$, its worldline might not be described by a Wilson line in a representation of $G$, but by a line inducing holonomy in some conjugacy class of $G$ around it \cite{Witten:1988hf,Elitzur:1989nr,Moore:1989yh,Gukov:2008sn}.
\item If $a$ is light near the transition, so are all its powers.  We choose one of them, $a$, and describe the transition as its proliferation.  For example, this chosen $a$ could be the lightest of them.  If the transition is triggered by a power of it, we should choose another $a$, whose corresponding field $\Phi$ has different couplings, leading to a potentially different transition theory and different $\cal T'$.
\end{itemize}

We will explore the transition theory, identify its dependencies, and determine its phase diagram.  In particular, we would like to describe a phase where $\Phi$ Higgses $G$ to a subgroup $G'$.  If this Higgs phase is gapped, it is described by the TQFT ${\cal T}'$.

Importantly, even though our TQFT can be non-Abelian, we will limit ourselves to the case where the proliferating anyons are Abelian.  This means that it generates a one-form global symmetry \cite{Gaiotto:2014kfa}. A special case of our discussion, where the proliferating anyons generate an anomaly-free one-form symmetry (i.e., the anyon is a boson), is closely related to gauging that symmetry (see Appendix \ref{sec:gauging-one-form}).  However, our discussion is more general and includes also anomalous one-form symmetries.  And as we will see, even when the symmetry is anomaly-free, the outcome is not unique.

We should mention certain situations that are not covered by our general discussion:
\begin{itemize}
\item We limit ourselves to bosonic $\cal T$.  Even then, the field theory describing the transition could include fermionic fields.\footnote{If the theory is bosonic, then the fermionic fields should be coupled to gauge fields in such a way that local gauge-invariant operators are bosonic. For a recent discussion of bosonic, fermionic, and electronic theories, see \cite{Cheng:2025ube}.}  In some cases, these fermionic theories are dual to bosonic theories (see e.g., \cite{Barkeshli:2012rja, Karch:2016sxi,seiberg2016duality}), and then some of them can be covered by our picture.  But the generic fermionic theory is outside our framework.
\item We assume that $\Phi$ is in a one-dimensional complex (or real) representation of $G$.  Our discussion does not apply to higher-dimensional representations.  For example, if the TQFT $\cal T$ is described by a non-Abelian $G$, the anyon $a$ can correspond to a multi-dimensional representation of $G$.  Then, the corresponding field $\Phi$ can also be multi-dimensional, and hence, this is not covered by our picture.   Many authors have studied Chern-Simons-Matter theories with matter fields in multi-dimensional representations.  For some recent examples in a context similar to ours, see e.g., \cite{Barkeshli:2012rja, MaQCD3, LeeQED3, song2024phase}. As we said above, in some cases, the TQFT can have a dual Chern-Simons description with another $G$ and another representation of $a$, such that $a$ is in a one-dimensional representation.  Then, our discussion applies, but the answers can differ from those with a multi-dimensional $\Phi$.
\item The anyon $a$ might be non-Abelian.  This means that it generates a non-invertible symmetry.  In that case, all the anyons in that symmetry also participate in the transition.  This case clearly falls outside the scope of our framework. See \cite{Cordova:2025eim} for a related study.
\item There can be multi-critical points where several different anyons become light and trigger the transition.  This situation is particularly interesting when these anyons are not relatively local, as in the recent work \cite{shi_analytic_2024,Dumitrescu:2026vre, Ji:2026yfj}.
\end{itemize}

In Section \ref{twoTQFTs}, we will present the two TQFTs $\cal T$ and $\cal T'$, and will discuss several relations between them.  In Section \ref{nontoptrant}, we will embed these TQFTs in the non-topological transition field theory. (See Figure \ref{threescales}.)  Here, we will exclude a special case where a certain integer $p$ vanishes.  Then, we will discuss the system's global symmetries and its dynamics.

Section \ref{piszero} will be devoted to the special case of $p=0$, where our transition field theory implements gauging a one-form symmetry.

In Section \ref{dual theory}, we will present a particle/vortex dual description of our transition theory.  It leads to the same two phases as the original theory, but this time, the opposite transition is a Higgs transition.

In Section \ref{enrichmentsec}, we will enrich $\cal T$ with a global $\U_A$ and explore the coupling of the background field $A$ to the full transition theory.  One interesting conclusion is that in the special case where the transition theory implements gauging a one-form symmetry, and the anyon $a$ carries fractional $\U_A$ charge, the enrichment is incompatible with our transition theory.

We will summarize our results and point out interesting directions for further study in Section \ref{conclusions}.

Finally, in Appendix \ref{sec:gauging-one-form}, we will review the process of gauging a one-form symmetry.

\section{A Tale of Two TQFTs}\label{twoTQFTs}

\subsection{$\cal T$}\label{calT}

We consider a bosonic TQFT $\cal T$, and we focus on an Abelian anyon $a\in {\cal T}$ of order $n$, i.e., $a^n={\bf 1}$ is the identity in $\cal T$.  The anyon $a$ generates a $\bZ_n^{(1)}$ symmetry with anomaly $p$.  This means that its spin is
\ie\label{hofa}
h(a)={p\over 2n}\bmod 1\,,
\fe
and since $\cal T$ is a bosonic theory, $pn\in 2\bZ$ \cite{Hsin:2018vcg}.  As far as the TQFT is concerned, $p \sim p+2n$.  However, we will view $p$ as an arbitrary integer.  

We will find it convenient to define
\ie\label{Ldef}
L=\gcd(n,p)\,.
\fe
 For $L=n$, i.e., $p=0\bmod n$, the anomaly vanishes.   For $p=0\bmod 2n$, this means that the symmetry can be gauged. (See Appendix \ref{sec:gauging-one-form}.) And for $p=n\bmod 2n$, the symmetry can be gauged, but the gauging produces a fermionic theory. For $L=1$,  the TQFT factorizes as ${\cal T}=\hat {\cal T}\boxtimes {\cal A}^{n,p}$ with ${\cal A}^{n,p}$ a canonical minimal theory with that symmetry and anomaly \cite{Moore:1988qv,Hsin:2018vcg}.

We will also be interested in the $\bZ_n$ gauge theory with a Dijkgraaf-Witten twist \cite{Dijkgraaf:1989pz} labeled by $pn$, ${\cal T}_{n,p}$.  We will describe it using the Lagrangian
\ie\label{CnpL}
{\cal L}_{{\cal T}_{n,p}}=-{np\over 4\pi} {b}d{b} +{n\over 2\pi} {b}dc\qquad , \qquad pn\in 2\bZ\,,
\fe
where $ b$ and $c$ are $\U$ gauge fields.   By redefining $c\rightarrow c+b$, we can shift $p$ by $2$ and therefore, for odd $n$, we can set $p=0$, and for even $n$, we can limit ourselves to $p=0,1$.  However, since later we will have a preferred definition of $c$, we do not use the freedom to change $p$.

The anyon $\exp\left(i\int(p{b} -c)\right)$ in the theory ${\cal T}_{n,p}$ generates a $\bZ_n^{(1)}$ symmetry with anomaly $-p$, as can be seen using
\ie\label{hanyonDW}
h\left(e^{i\int(p{b} -c)}\right)=-{p\over 2n}\bmod 1\,.
\fe

We will show that the two TQFTs, ${\cal T}$ and ${\cal T}_{n,p}$ satisfy \cite{Hsin:2016blu,Cordova:2017vab,Hsin:2018vcg}
\ie\label{CCCrel}
{\cal T}\cong{{\cal T}\boxtimes{\cal T}_{n,p} \over \bZ_n}= {{\cal T}\boxtimes{\cal T}_{n,p} \over (a, p,-1)}\,,
\fe
where the quotient means gauging the $\bZ_n^{(1)}$ one-form symmetry generated by $(a, p,-1)$, which corresponds to the anyon $a\exp\left(i\int(p{b} -c)\right)$. (See Appendix \ref{sec:gauging-one-form}.)  This is a valid quotient, as can be seen by using \eqref{hofa} and \eqref{hanyonDW}.  

Before deriving \eqref{CCCrel}, we consider the effect of the quotient \eqref{CCCrel} on the fields $ b$ and $c$ in  ${\cal T}\boxtimes{\cal T}_{n,p}$.  The equations of motion of $ b$ and $c$ in the presence of the line defect $a\exp\left(i\int_{{\cal C}_1}(p{b} -c)\right)$ state that $c$ is smooth and $d{ b}={2\pi\over n}\delta({\cal C}_1)$.  The quotient \eqref{CCCrel} amounts to summing over such line insertions.  (See Appendix \ref{sec:gauging-one-form}.)  Hence, $c$ is still a standard $\U$ gauge field.  However, in the quotient, $b$ becomes a twisted gauge field, which we denote as $\check  b$, whose fluxes have nonstandard quantization
\ie\label{fractionalb}
\int_{{\cal C}_2} d{\check  b}\in {2\pi\over n}\bZ\,,
\fe
with the fractional part correlated with the modes in $\cal T$.

In more detail, the quotient in \eqref{CCCrel} can be implemented by coupling the field theory in the numerator to a dynamical two-form $\bZ_n^{(1)}$ gauge field ${\mathbf g}^{(2)}$, which is a 2-cocycle, i.e., $\int_{{\cal C}_2}{\mathbf g}^{(2)}$ is an integer modulo $n$.\footnote{We will use the fonts ${\mathbf c}$, ${\mathbf g}$, etc. to denote discrete fields and more standard fonts to denote continuous fields.}  Then,
\ie\label{frakbcyc}
{1\over 2\pi}\int_{{\cal C}_2}d{\check  b}={1\over 2\pi}\int_{{\cal C}_2}d b +{1\over n}\int_{{\cal C}_2} {\mathbf g}^{(2)}\in {1\over n}\int_{{\cal C}_2} {\mathbf g}^{(2)}+\bZ \,.
\fe
Here, $b$ in the middle expression is an ordinary $\U$ gauge field.\footnote{Let us compare this equation to more familiar situations.  Consider a scalar field $h\sim h+2\pi$ with a $\U^{(0)}$ gauge symmetry $h\to h-\lambda$, which is coupled to a $\U^{(0)}$ gauge field $a^{(1)}$ with $a^{(1)}\to a^{(1)}+d\lambda$ via $dh+a^{(1)}$.  When $da^{(1)}=0$, it is common to write it as $d\check {h}=dh+a^{(1)}$, and to refer to $\check  h$ as a twisted scalar.  Similarly, a $\U^{(0)}$ gauge field $a^{(1)}$ with a $\U^{(1)}$ gauge symmetry $a^{(1)}\to a^{(1)}-\lambda^{(1)}$ couples to a gauge field $b^{(2)}$ with $b^{(2)}\to b^{(2)}+d\lambda^{(1)}$ via $da^{(1)}+b^{(2)}$.  When $db^{(2)}=0$, we can write it as $d\check {a}^{(1)}=da^{(1)}+b^{(2)}$, and refer to $\check  {a}^{(1)}$ as a twisted gauge field.  Equation \eqref{frakbcyc} is similar, except that the one-form symmetry is discrete, and hence, its gauge field ${\mathbf g}^{(2)}$ is a cocycle.}  This means that\footnote{Note our notation.  We started with a $\U$ gauge field $b$.  We twisted it to $\check  b$, and now we define a new standard $\U$ gauge field $\frak b$.  Hopefully, the reader will not find this notation too confusing. If they do, we apologize for that.\label{differentbs}} 
\ie\label{btilded}
{\frak b}=n\check {b}
\fe
is a standardly normalized $\U$ gauge field and it couples to the modes of $\cal T$ in the numerator of \eqref{CCCrel} only through
\ie\label{fluxcoup}
\left({1\over 2\pi}\int_{{\cal C}_2}d{\frak b} \right) \bmod n=\int_{{\cal C}_2} {\mathbf g}^{(2)}\,,
\fe
which we denote as
\ie\label{cocyclno}
{\mathbf g}^{(2)}=\left[{d{\frak b}\over 2\pi}\right]_n\,.
\fe

 It is important that $\frak b$ is $\bZ_n^{(1)}$ gauge invariant and it contains all the $\bZ_n^{(1)}$ gauge invariant information in ${\mathbf g}^{(2)}$ and $\check  b$.  In particular, it determines  ${\mathbf g}^{(2)}$ via \eqref{cocyclno}, and $\check  b$ is determined via \eqref{btilded}, up to a $\bZ_n^{(1)}$ gauge transformation given by shifting it by a $\bZ_n$ gauge field.

 One way to derive \eqref{CCCrel} is as follows.  We label the anyons in the numerators as $(s, w_{b}, w_c)$ with $s\in {\cal T}$ and the other labels correspond to the anyons $\exp\left(i\int(w_{b}{b} +w_cc)\right)\in {\cal T}_{n,p} $ (with $w_b,w_c$ integers modulo $n$).\footnote{Here, we do not use $\check  b$, because these are fields in the numerator before taking the quotient.}  First, using the identification by $(a, p,-1)$, we can set $w_c=0$. (Since this identification does not have any fixed points, no anyon is split.)  Next, every anyon $s\in \cal T$ transforms under $a$ as $e^{2\pi i n_s\over n}$, with integer $n_s$.\footnote{Here, only $n_s\bmod n$ is important.  But for later convenience, we will treat $n_s$ as an integer.\label{nsisinteg}} To make it braid trivially with $a\exp\left(i\int(p{b} -c)\right)$, we should have $w_{b}=n_s\bmod n$. Therefore, any anyon $s\in {\cal T}$ is mapped in the quotient to $(s,n_s\bmod n,0)$, and this map is onto.  Then, it is easy to check that the anyons $(s,n_s\bmod n,0) $ satisfy the same fusion as $s\in {\cal T}$, thus establishing \eqref{CCCrel}.  Specifically, the anyon $a\in {\cal T}$ is mapped to $(a,p,0)\sim ({\bf 1},0,1)$, i.e., to $\exp\left(i\int c\right)$.

In Section \ref{interpretationI}, we will provide another derivation of \eqref{CCCrel} based on integrating out $c$.

We see that the presentation of $\cal T$ as ${{\cal T}\boxtimes{\cal T}_{n,p} \over \bZ_n}$ focuses on the anyon $a\in {\cal T}$, and describes it simply as $\exp\left(i\int c\right)$.

\subsection{${\cal T}'$}\label{Tprimed}

For the moment, let us assume that $p\ne 0$.  We will return to $p=0$ below.

We repeat the previous discussion, replacing \eqref{CnpL} with
\ie\label{LUnp}
{\cal L}_{\U_{-np}}=-{np\over 4\pi} {b}d{b} \qquad , \qquad pn\in 2\bZ\,.
\fe
Here, the line $\exp\left({ip\int {b}}\right)$ generates a $\bZ_n^{(1)}$ symmetry with anomaly $-p$.  Therefore, instead of \eqref{CCCrel}, we can consider
\ie\label{Cprimed}
{\cal T}'={{\cal T}\boxtimes \U_{-pn}\over \bZ_n} = {{\cal T}\boxtimes \U_{-pn}\over (a,p)}\,.
\fe
It is easy to verify that this is a valid quotient, i.e., the denominator is an anomaly-free one-form symmetry.\footnote{The case with $L=\gcd(n,p)=1$ is special.  Here, ${\cal T}$ and $\U_{-np}$ factorize as
\ie
&{\cal T}=\hat{\cal T}\boxtimes {\cal A}^{n,p}\\
&\U_{-np}\equiv{\cal A}^{n,-p} \boxtimes {\cal A}^{p,-n}\,,
\fe
with ${\cal A}^{n,p}$ a canonical minimal theory with that symmetry and anomaly \cite{Moore:1988qv,Hsin:2018vcg}, and we took, for simplicity, $p>0$.  Then,
\ie
{\cal T}'={(\hat{\cal T}\boxtimes {\cal A}^{n,p})\boxtimes ({\cal A}^{n,-p} \boxtimes {\cal A}^{p,-n})\over \bZ_n} =\hat{\cal T}\boxtimes{ {\cal A}^{n,p}\boxtimes {\cal A}^{n,-p}\over \bZ_n} \boxtimes {\cal A}^{p,-n} = \hat{\cal T}\boxtimes{\cal A}^{p,-n}\,.
\fe
So the map from $\cal T$ to ${\cal T}'$, simply replaces the ${\cal A}^{n,p}$ factor with ${\cal A}^{p,-n}$. Note that the map depends on $p$, not only $p\bmod 2n$, through the dependence of ${\cal A}^{p,-n}$ on $p$. \label{Lisonef}}

The construction of $\cal T'$ in \eqref{LUnp} is a special case of the ``stack and condense'' procedure, used in different contexts by various authors, including \cite{LanHierarchy2017,Lan:2019gxv, Hsin:2018vcg, Cheng:2022nds, zhang2024hierarchy,Shi2025}.   In our special case, we stack $\U_{-np}$.  As we will see below, this particular construction allows us to embed both $\cal T$ and $\cal T'$ in the same simple transition field theory.

Again, we can describe this quotient by turning the gauge field $b$ in \eqref{LUnp} into a twisted gauge field $\check  b$, and then using ${\frak b}=n\check  {b}$, and ${1\over 2\pi}\int_{{\cal C}_2}d{\frak b} \bmod n=\int_{{\cal C}_2} {\mathbf g}^{(2)}$ as in \eqref{fractionalb} - \eqref{fluxcoup}.

\subsubsection{The zero-form symmetry of ${\cal T}'$}\label{symofTp}

Below, we will study some subtle zero-form symmetries of TQFTs and non-topological theories.  As a warm-up, which will highlight some of the notions we will use, we discuss the zero-form symmetries of $\cal T'$, which arise from gauging the $\bZ_n^{(1)}$ in \eqref{Cprimed}.  (See Appendix \ref{app:dualsymmetry}.)

TQFTs like $\cal T$ or ${\cal T}'$ do not have local operators.  Yet, they can still have a global zero-form symmetry, which acts on line operators and their junctions \cite{etingof2010fusion, barkeshli_symmetry_2019}.  Often the symmetry permutes lines, although it is not always the case (e.g. \cite{Davydov2014Bogomolov, Kobayashi:2025ykb}).\footnote{Global zero-form symmetries that are used to enrich the theory using its one-form symmetry are not genuine symmetries.}

Naively, the $\bZ_n^{(1)}$ gauging, associated with the quotient in \eqref{Cprimed}, leads to a dual $\bZ_n^{(0)}$ generated by
\ie\label{Zerogene}
\exp\left({2\pi i\over n} \int_{{\cal C}_2} {\mathbf g}^{(2)}\right)=\exp\left({i\over n} \int_{{\cal C}_2} d{\frak b}\right)\,.
\fe
However, this zero-form symmetry acts trivially on the TQFT $\cal T'$.  A quick way to see it uses the equation of motion of $\check  b$ to learn that $d\check {b}=0$ and therefore,
\ie\label{fluxrestr}
 d{\frak b}=0\,.
 \fe
(Recall that in this discussion, we limit ourselves to nonzero $p$. Below, we will return to the case of $p=0$.)    We stress that the flux restriction \eqref{fluxrestr} is not a restriction on the configurations in the functional integral; it is valid only on-shell.

A more detailed derivation of the trivial action of \eqref{Zerogene} is as follows. A faithful action is related to the splitting of anyons by the quotient in \eqref{Cprimed}.  Let us examine the quotient in more detail.  An anyon $(s,w_b)\in {\cal T}\boxtimes \U_{-pn}$ is included in the quotient only when it braids trivially with $(a,p)$, i.e., when $n_s=w_b\bmod n$.  (Compare with the discussion following \eqref{cocyclno}, and note that, unlike the case there, here $w_b$ is defined modulo $np$.)  These anyons are subject to the identifications $(s,w_b)\sim (as,w_b+p)$.  Since $w_b$ is defined modulo $np$, this identification does not have fixed points.  And consequently, the dual symmetry generated by \eqref{Zerogene} acts trivially on the TQFT.

 We conclude that the gauging in \eqref{Cprimed} does not produce a new zero-form symmetry.  Having said that, we should stress that typically the original theory $\cal T$, and therefore also  $\cal T'$, can have faithfully-acting zero-form symmetries.

\subsection{The relation between ${\cal T}$ and ${\cal T}'$}\label{TTprel}

We have discussed two TQFTs, $\cal T$ and $\cal T'$.  Here, we will elaborate on the relation between them.

The presentations \eqref{CCCrel} and \eqref{Cprimed} of these two theories point to the duality relation \cite{Hsin:2016blu,Cordova:2017vab,Hsin:2018vcg}
\ie\label{Cpp}
{\cal L}_{{\cal T}'}  +{1\over 2\pi}{\frak b}dc\quad\longleftrightarrow \quad {\cal L}_{{\cal T}}\,.
\fe
In more detail,  ${\cal L}_{\U_{-np}}$ of \eqref{LUnp} together with the added term ${1\over 2\pi}{\frak b}dc$ are the same as ${\cal L}_{{\cal T}_{n,p}}$ of \eqref{CnpL}.  This suggests that the TQFT of \eqref{Cpp} can be described by replacing the $\U_{-np}$ factor in the numerator of \eqref{Cprimed} with ${\cal T}_{n,p}$ of \eqref{CnpL}.  This is almost correct.  To make it completely correct, we also have to replace the denominator $(a,p)$ in \eqref{Cprimed} with $(a,p,-1)$ in \eqref{CCCrel}.

Below, we will present three interpretations of the map from $\cal T$ to $\cal T'$ and the relation \eqref{Cpp}. In Section \ref{twoexamples}, we will demonstrate these interpretations of the map from $\cal T$ to ${\cal T}'$ in two examples.

We should point out that even though ${\cal L}_{{\cal T}'}$ describes a meaningful TQFT only for nonzero $p$, the relation \eqref{Cpp}  is valid for all $p$, including $p=0$.  In particular,  ${\cal L}_{{\cal T}'}  +{1\over 2\pi}{\frak b}dc$ describes a consistent theory for all $p$.

\subsubsection{Derivation of \eqref{Cpp} by integrating out $c$}\label{interpretationI}

 One way to interpret \eqref{Cpp} is to note that the gauge field $c$  sets ${\frak b}$ to zero, or more precisely, to a pure gauge.  This means that both the twisted gauge field $\check  b$ and $\mathbf g^{(2)}$ can be set to zero. (Recall that after \eqref{cocyclno} we stated that all the $\bZ_n^{(1)}$ gauge invariant information in $\check  b$ and $\mathbf g^{(2)}$ is included in $\frak b$, which we now set to zero.)  Therefore, the coupling to $c$ removes the factor of $\U_{-np}$ in the numerator of $\cal T'$, and removes the quotient because the dynamical field $\mathbf g^{(2)}$ vanishes.  We end up with $\cal T$.

\subsubsection{Interpretation of \eqref{Cpp} as coupling $\cal T$ to a $\U$ gauge field}\label{Cppgau}

Another interpretation of the relation between $\cal T$ and $\cal T'$ follows \cite{Cheng:2022nds}.   Let us couple $\cal T$ to a $\U_X$ background gauge field $X$ by adding to the left-hand side of \eqref{Cpp} the terms\footnote{We follow the convention that classical gauge fields are denoted by uppercase letters, and dynamical gauge fields are denoted by lowercase letters.  When we turn a classical field into a dynamical field, we denote it by the lowercase version of the same letter.\label{uppercasec}}
\ie\label{Xccoupling}
{1\over 2\pi}Xdc+{N\over 2\pi}XdX \qquad, \qquad N\in \bZ\,.
\fe
This coupling of $X$ to $\cal T$ means that we enriched $\cal T$ with $\U_X$.  Specifically, we have already noted that in the presentation of $\cal T$ as ${{\cal T}\boxtimes{\cal T}_{n,p} \over \bZ_n}$, the anyon $a$ is described by $\exp\left(i\int c\right)$.  Then, the first term in \eqref{Xccoupling} means that $a$ is the vison for $\U_X$.  Equivalently, $a$ generates $\bZ_n^{(1)}$, and using the notation in \eqref{cocyclno}, the $\bZ_n$-valued 2-cocycle $\left[{dX\over 2\pi}\right]_n$ is its background gauge field. Also, it is clear that the second term in \eqref{Xccoupling} represents a Symmetry-Protected Topological (SPT) phase for this $\U_X$.

Next, we turn the background field $X$ into a dynamical $\U$ gauge field $x$.  Now, $c$ sets $x=-{\frak b}$ (up to a gauge transformation). And we end up with $\cal T'$ with $p\to p- 2nN$.

The conclusion is that starting with $\cal T$, we use its $\bZ_n^{(1)}$ symmetry to enrich it with $\U_X$.  We also use the freedom to add an SPT for $\U_X$.  Then, we make the background $X$ dynamical to find $\cal T'$, where the freedom in adding the SPT translates into shifting $p$ by an integer multiple of $2n$ \cite{Cheng:2022nds}.

\subsubsection{Interpretation of \eqref{Cpp} as a hierarchy construction}\label{heirarchycon}

A well-known construction of new Fractional Quantum Hall (FQH) states is the (Abelian) hierarchy construction.  In its original form, starting from a given FQH wavefunction and an Abelian anyon $a$, one writes a new wavefunction by forming a Laughlin state out of a finite density of the chosen anyon. This procedure can then be iterated to create increasingly more complicated states. 

The construction was first formulated for Abelian FQH states \cite{HaldaneHierarchy, HalperinHierarchy}, and the corresponding Abelian TQFT description was developed in \cite{Blok:1990mc, Blok:1990an, ReadHierarchy1990,wen1992classification}.  It was later generalized to certain non-Abelian states \cite{BondersonSlingerland}. An algebraic description of this construction, expressed as a map between two TQFTs, was given in \cite{LanHierarchy2017,Lan:2019gxv}. It was then recognized in \cite{Cheng:2022nds} that this mapping is equivalent to \eqref{Cprimed}.

To see why this is the case, let us first treat $c$ as a classical background field, and following footnote \ref{uppercasec}, denote it by $C$.  We couple it to $\U_{2N}$ with the Lagrangian ${2N\over 4\pi} xdx +{1\over 2\pi}x dC$.  Here, the Wilson line $e^{i\int C}$ represents the worldline of a transparent particle.  In the FQH framework, this transparent line is the electron (except that our system is bosonic) and $x$ is the statistical gauge field for a $\nu={1\over 2N}$ state.  

Next, we consider $\cal T$ with a dynamical $c$.  Here, $e^{i\int c}$ represents the anyon $a$.  Coupling this to the system with $x$, we identify $a$ as an ``electron'' leading to a FQH state with $\nu={1\over 2N}$.  The result is precisely \eqref{Xccoupling} (after making $X$ dynamical).

Finally, we note that there are also further generalizations of hierarchy constructions in which non-Abelian anyons form many-body topological states \cite{ levin2009collective,HermannsPRL2010,Hansson2017quantumhall,zhang2024hierarchy}.  This is beyond the scope of our discussion.

\subsubsection{The inverse map with the same procedure}\label{inversemap}

Above, we studied a map from $\cal T$ to $\cal T'$.  It depends on a choice of an Abelian anyon $a\in {\cal T}$ of order $n$ and an integer $p$.  Here, we will apply the same procedure starting with $\cal T'$ and mapping it back to $\cal T$.

Consider nonzero $p$ and assume, for simplicity, that $p$ is positive.  We  choose $\tilde a=({\bf 1},n)\in {\cal T}'$, corresponding to $e^{i\int \frak b}$, whose order is $\tilde n=p$ and its spin is $-{n\over 2 p}\bmod 1$. We also choose $\tilde p=-n$.  Then, applying our map to $\cal T'$ 
\ie\label{doublemap}
&{\cal T}'= {{\cal T}\boxtimes \U_{-pn}\over (a,p)} \longrightarrow {\cal T}''\\
&{\cal T}''= {{\cal T'}\boxtimes \U_{pn}\over (\tilde a,-n)}= {{\cal T}\boxtimes \U_{-pn}\boxtimes \U_{pn}\over (a,p, 0)\times({\bf 1},n,-n)}\,.
\fe
In order to unpack this expression, we write the Lagrangian for $\U_{-pn}\boxtimes \U_{pn}$ as $-{np\over 4\pi}( b^1db^1- b^2db^2)$.  Then, the quotient by $(n,-n)$ can be performed by turning $b^1$ and $b^2$ into twisted fields $\check b^1$ and $\check b^2$ with appropriate flux quantization such that $e^{in\int(\check b^1-\check b^2)}$ is transparent.  These twisted fields can be expressed in terms of  standard $\U$ gauge fields $b$ and $c$ as $\check b^1=b-{1\over p}c$ and $\check b^2=-{1\over p}c$. This turns the Lagrangian for $\check b^1$ and $\check b^2$ into
\ie\label{}
{\cal L}_{{\cal T}_{n,p}}=-{np\over 4\pi} {b}d{b} +{n\over 2\pi} {b}dc
\fe
of \eqref{CnpL}. And  $e^{in\int(\check b^1-\check b^2)}=e^{in\int b}$ is indeed transparent.  The final quotient in \eqref{doublemap} is by $a e^{ip\int \check b^1}=ae^{i\int( pb-c)}$.  So we conclude that
\ie
{\cal T}''={{\cal T}\boxtimes {\cal T}_{n,p} \over (a,p, -1)}=\cal T\,,
\fe
where we used \eqref{CCCrel}.

In conclusion, the original map was from $\cal T$ to $\cal T'$.  It depended on $a\in {\cal T}$ of order $n$ and $p$.  Repeating this map from $\cal T'$ with $\tilde a=({\bf 1},n)\in {\cal T}'$ of order $\tilde n=p$ and $\tilde p=-n$ leads back to $\cal T$.  (Above, we derived it for positive $p$.  A similar result is valid for negative $p$.)  Clearly, this generalizes the statement in footnote \ref{Lisonef} to the case with $L\ne 1$.

\section{The non-topological transition field theory}\label{nontoptrant}

We would like to describe the most general transition from a gapped phase whose low-energy theory is the TQFT $\cal T$, in which a particle with the quantum numbers of $a$ becomes light and triggers the transition.  We assume that this particle can be described by a single complex scalar field $\Phi$, rather than by a large multiplet.  We also assume that the dynamics is Lorentz-invariant.\footnote{Even though we assume relativistic invariance, we point out that there are non-relativistic systems with similar physics.}

For that, we write the Lagrangian\footnote{Throughout this note, we abuse the notation by suppressing the wedge product of forms, and by adding differential forms and scalar terms.  We also suppress coefficients of order one that do not affect our discussion.  The ellipses represent higher-order potential terms for $\Phi$.
\label{absuse}}
\ie\label{Higgstheory}
{\cal L}_{{\cal T}'}  +{1\over 2\pi} {\frak b}dc +|D_c\Phi|^2 -\mu^2|\Phi|^2 + \cdots
\fe
The first two terms in the Lagrangian describe the original TQFT $\cal T$ \eqref{Cpp}.  The field $\Phi$ couples to this TQFT through its coupling to $c$.\footnote{Our conventions are that a scalar field $\chi$ with $\U_B$ charge $1$ couples to the gauge field $B$ using the covariant derivative  $D_B\chi=(d -i B)\chi$.  Then, $\chi({\cal P}) \exp\left({i\int_{\cal P}^{{\cal P}'} B}\right)\bar\chi({\cal P}')$ is gauge invariant.  In these conventions, the $\U_B$ current is $J_B=-{\delta {\cal L}\over \delta B}$.  Therefore, a kinetic term like $|D_B\chi|^2$ contributes to $J_B$
\ie
\mathscr{J}(\chi)=\star i\chi \overset\leftrightarrow{D} \bar\chi=\star i(\chi {D}\bar\chi-\bar\chi D\chi)\,,
\fe
where $D$ is the appropriate covariant derivative in each term. And a Chern-Simons term, ${k\over 2\pi} B dC$ contributes $J_B=- {k\over 2\pi} dC $. Here, $\star $ is the Hodge dual.  Note that we define the current as a differential form.  People unfamiliar with this notation can simply consider the Hodge dual of our expressions.    \label{convensionsf}} This means that $\Phi$ is a source of the anyon described by $\exp\left(i\int c\right)$, which we identified as the anyon $a$.  Therefore, intuitively, the field $\Phi$ creates our preferred anyon $a$.

\subsection{Local dynamics}\label{localdynamics}

Before we delve into a detailed analysis of the theory, we note that locally it is simplified.  Locally, the modes in $\cal T$ and their correlations with the flux of $\check  b$ are not important.  Therefore, our theory is the same as
\ie
-{np\over 4\pi} \check {b}d\check {b} +{n\over 2\pi} \check {b}dc +|D_c\Phi|^2 -\mu^2|\Phi|^2 + \cdots
\fe
Also, locally, $\check  b$ and $c$ and therefore also ${\frak b}=n\check  {b}$ can be treated as $\bR$ gauge fields.  Then, we can integrate out $\check  b$ and find
\ie\label{dualtheory}
{n\over 4\pi p} cdc +|D_c\Phi|^2 -\mu^2|\Phi|^2 + \cdots
\fe
The fractional Chern-Simons coefficient is consistent because we treat this theory locally, and then $c$ is an $\bR$ gauge field.

We conclude that locally, our theory depends only on one rational number, $n\over p$.  For the special case of $p=0$, the divergence in the first term means that locally, we can set $c=0$ and $\Phi$ does not couple to any gauge field.  We will elaborate on this in Section \ref{localdynpzero}.

\subsection{The global symmetries}\label{theglobal s}

\subsubsection{The one-form symmetry}

 Here we discuss the one-form symmetry of the system.  We limit ourselves to the invertible symmetry.  In the TQFT $\cal T$, these are the Abelian anyons.

The coupling of the TQFT $\cal T$ to $\Phi$ breaks explicitly some of its one-form global symmetry.

In order to determine the unbroken symmetry, we use the presentation of $\cal T$ with the fields ${\frak b}$ and $c$ in \eqref{Cpp} and recall that every symmetry operator of $\cal T$ is an anyon line.  One of the effects of the coupling to the scalar field $\Phi$ is that the Wilson line of $c$ can end on $\Phi$.  Therefore, every anyon line in $\cal T$ that braids non-trivially with  $\exp\left(i\int c\right)$ cannot be a symmetry line.\footnote{Note that a symmetry line can end at a point, but a charged line cannot end at a point.  In our case, the lines $e^{im \int c}$ can end on $\Phi^m$.  Therefore, the lines $e^{im\int c}$ can be symmetry lines.  However, they cannot be charged under any symmetry line.  Therefore, lines that braid nontrivially with $e^{i\int c}$ including $e^{im\int c}$ with ${m p\over n}\notin \bZ$ cannot be symmetry lines.}

We conclude that the unbroken one-form symmetry of our field theory is generated by all the lines of $\cal T$ that braid trivially with the anyon $a$. This reasoning applies both to invertible and non-invertible symmetries.

 In particular, the $\bZ_n^{(1)}$ symmetry generated by the anyon $a$ is broken to  $\bZ_{L}^{(1)}$ generated by $a^{n\over L}$ corresponding to the Wilson line $\exp\left({in\over L} \int c\right)$. (Recall that  $L=\gcd(n,p)$.) Given that the anomaly of the original $\bZ_n^{(1)}$ was $p\bmod 2n$, the anomaly of this $\bZ_{L}^{(1)}$ is ${pn\over L}\bmod 2L$. Since $np$ is even, this anomaly vanishes unless both $n\over L$ and $p\over L$ are odd, and then it is $L\bmod 2L $.

\subsubsection{The zero-form symmetry}\label{zeroforms}

The original TQFT has no faithfully acting continuous zero-form symmetries.\footnote{The Chern-Simons description of the TQFT can include $\U$ gauge fields, whose field strengths are conserved because of the Bianchi identity.  However, all these currents vanish on-shell.  The equations of motion set them to zero.  In Section \ref{enrichmentsec}, we will enrich the TQFT with a $\U_A$ global symmetry, but it will still not act faithfully on it.}  This is not the case in our transition theory.  In the absence of kinetic terms for the gauge fields, there is one continuous zero-form global symmetry, which we denote $\U_{\cal B}$.

Let us find its current $J_{\cal B}$.  As a first attempt, we try
\ie
\mathscr{J}(\Phi) =\star i\Phi\overset\leftrightarrow{D} \bar\Phi\,,
\fe
under which $\Phi $ has charge $1$. (See footnote \ref{convensionsf}.)  This is not a gauge current because the equation of motion of $c$, which sets the gauge current of $c$ to zero, is ${1\over 2\pi}  d{\frak b}=\mathscr{J}(\Phi)$.  Therefore, this attempt to find $J_{\cal B}$ can be written as
\ie
\mathscr{J}(\Phi)= {1\over 2\pi}  d{\frak b}\,.
\fe
In its form on the left-hand side, it looks like it acts only on $\Phi$, while its form on the right-hand side shows that it can be taken to be a magnetic flux.  These two interpretations are the same on-shell.

$\mathscr{J}(\Phi)$ is a conserved current, and it leads to quantized charges.  However, the corresponding symmetry might not act faithfully, so the true conserved current $J_{\cal B}$ could be a fraction of $\mathscr{J}(\Phi)$.  This could happen if the fluxes of ${\frak b}$ are not arbitrary integers.  (Compare with the discussion in Section \ref{symofTp}.)

One restriction on the fluxes of ${\frak b}$ arises from its own equation of motion, equivalently, from its gauge symmetry, which sets $pd{\frak b}=ndc$.  Using the off-shell flux quantization ${1\over 2\pi}\int_{{\cal C}_2}d{\frak b}, {1\over 2\pi}\int_{{\cal C}_2}dc \in \bZ$, the on-shell fluxes must satisfy
\ie\label{restfrombt}
{1\over 2\pi}\int_{{\cal C}_2}d{\frak b}\in {n\over L}\bZ\,.
\fe

Since the fluxes of ${\frak b}$ are restricted, our second attempt at finding the current $J_{\cal B}$ is
\ie\label{secondatt}
{L\over n}\mathscr{J}(\Phi)= {L\over 2\pi n}  d{\frak b}={m_b\over 2\pi}  d{\frak b}+{m_c\over 2\pi}  d{c}\,,
\fe
 with $m_{\frak b}$ and $m_c$ integer solutions of $m_bn+m_cp=L$.  Again, the first expression shows that the charge can be taken to act only on $\Phi$.\footnote{The fact that the charge of $\Phi$ using ${L\over n}\mathscr{J}(\Phi)$ is fractional is consistent because $\Phi$ is not gauge invariant.  Our normalizations are such that all gauge-invariant local operators have integer charges. A similar situation arises in QCD, where the baryon number of a quark is $1\over 3$, but all gauge-invariant operators have integer baryon number.\label{fractionachargeB}}  The last expression is related to the first two expressions using the equations of motion.  In this form, ${L\over n}\mathscr{J}(\Phi)$ leads to integer charges even off-shell.

Again, ${L\over n}\mathscr{J}(\Phi)$ is conserved, and it leads to quantized charges, but it might still not act faithfully. This happens when the equations of motion of the dynamical fields in $\cal T$ restrict the fluxes of ${\frak b}$ beyond \eqref{restfrombt}.

A simple way to determine the restrictions of $\cal T$ on the fluxes of ${\frak b}$ is the following.  We tentatively think of ${\frak b}$ as a background field. This means that the TQFT is decoupled from the rest of the theory. Then, the coupling of ${\frak b}$ to $\cal T$ is through its $\bZ_n^{(1)}$ symmetry, which is generated by $a$.  Specifically, we use ${\frak b}$ to form a $\bZ_n$-valued 2-cocycle ${\mathbf g}^{(2)}$
(see \eqref{frakbcyc}) and couple it to $\cal T$ as a background field for its $\bZ_n^{(1)}$ symmetry.  This background field can be interpreted as defects of the anyon $a$.  The number of such defects piercing ${\cal C}_2$ is $\int_{{\cal C}_2} {d\frak{b}\over 2\pi} \bmod n=\int_{{\cal C}_2} {\mathbf g}^{(2)} \bmod n$.

Now, consider the Hilbert space of the theory on a spatial compact Riemann surface ${\cal C}_2$ with such a background field, but without any additional defects.

Let us start with ${\cal C}_2=S^2$.  Here, the fact that the Hilbert space with any nontrivial defect is empty means that the number of $a$ insertions should be a multiple of $n$, leading to the on-shell flux restriction
\ie\label{sphererestriction}
{1\over 2\pi}\int _{S^2} d{\frak b} \in n\bZ\,.
\fe
As we move to ${\cal C}_2$ with higher genus, depending on $\cal T$, there can be states in the Hilbert space with insertions of $a^m$ with $m$ a factor of $n$, $m|n$, leading to a weaker restriction
\ie\label{smallerflux}
{1\over 2\pi}\int _{{\cal C}_2} d{\frak b} \in m\bZ\qquad, \qquad m|n\,.
\fe

With this information, we make ${\frak b}$ dynamical and combine
\eqref{smallerflux} with \eqref{restfrombt}, which was determined by the equation of motion of ${\frak b}$.  We conclude
\ie\label{restfrombg}
{1\over 2\pi}\int_{{\cal C}_2}d{\frak b}\in {n\over \ell}\bZ\qquad, \qquad \ell|L\,,
\fe
where $\ell$ depends on the details of $\cal T$.

Hence, the properly-quantized $\U_{\cal B}$ current is
\ie\label{JBell}
J_{\cal B}={\ell\over n}\mathscr{J}(\Phi)= {\ell\over 2\pi n}  d{\frak b}\qquad , \qquad \ell|L\,.
\fe

Interestingly, because of \eqref{sphererestriction}, its subgroup $\bZ_\ell^{\cal B}\subset \U_{\cal B}$, which is generated by $\exp\left({2\pi i\over \ell}\int_{{\cal C}_2} J_{\cal B}\right)$,  does not act on the sphere without insertions.  Consequently, it does not act faithfully on local operators.  Only the quotient $\U_{\cal B}\over \bZ_\ell^{\cal B}$ acts on them.  However, this $\bZ_\ell^{\cal B}$ is a faithful symmetry of the theory, as it can act on line operators. 

The skeptical reader might wonder whether this discussion of $\ell|L$ is really needed.  In Section \ref{twoexamples}, we will present examples with the two extreme cases, $\ell=1$ and $\ell=L$ and will study the consequences of the value of $\ell$.

An important local operator is
\ie\label{monopoleo}
{\cal O}= \Phi^{n}{\cal M}_{\frak b}^n{\cal M}_c^{p}\,,
\fe
where ${\cal M}_{\frak b}$ and ${\cal M}_ c$ are monopole operators with $2\pi$ flux for ${\frak b}$ and $c$ respectively.  It is easy to check, using the conventions in footnote \ref{convensionsf}, that this operator is gauge invariant and satisfies the flux restriction \eqref{sphererestriction}.  It carries charge $\ell$ under the current $J_{\cal B}$ of \eqref{JBell}.  This is the minimal charge for local operators because only $\U_{\cal B}\over \bZ_\ell^{\cal B}$  acts on them.

We would like to make several points regarding the operator \eqref{monopoleo}.
\begin{itemize} 
\item Even though ${\frak b}=n\check  {b}$ and $\check  b$ couples to the modes in $\cal T$, $\cal O$ does not couple directly to $\cal T$.  The reason is that the factor of ${\cal M}_{\frak b}^n$ in $\cal O$ means that the flux of ${\frak b}$ around it is a multiple of $n$, and hence, the period of ${\mathbf g}^{(2)}$ around $\cal O$ is trivial. (See \eqref{frakbcyc}.)  Consequently, there is no direct coupling between the modes of $\cal T$ and $\cal O$.  

\item One might be tempted to consider a more basic operator
\ie\label{monopoleopp}
{\cal O}'= \Phi^{n\over \ell}{\cal M}_{\frak b}^{n\over \ell} {\cal M}_c^{p\over \ell}\,,
\fe
which carries charge 1 under $J_{\cal B}$ of \eqref{JBell}.  It is also gauge invariant and satisfies the flux restriction \eqref{restfrombg}.  However, since it does not satisfy the flux restriction on the sphere \eqref{sphererestriction}, this operator can be supported only on a line (or a line with junctions), which is surrounded by a  Riemann surface with positive genus.  Hence, ${\cal O}'$ of \eqref{monopoleopp} is not a local (point) operator.  This is consistent with our conclusion above that $\U_{\cal B}$ acts faithfully on the theory, but its $\bZ_\ell^{\cal B}$ subgroup does not act faithfully on local operators.  It does act faithfully on the line operator ${\cal O}'$.   

\item 
The form of the operator $\cal O$ \eqref{monopoleo} can be understood intuitively as follows.  As we commented above, the field $\Phi$ creates the anyon $a$.  Therefore, $\Phi^n$ creates the anyon $a^n={\bf 1}$, which is trivial in the TQFT.  In our Chern-Simons description of $\cal T$ with the fields $\frak b$ and $c$, $a^n={\bf 1}$ is the monopole operator in \eqref{monopoleo}.

\item 
The operator \eqref{monopoleo} has an interesting property.  The factor of ${\cal M}_c^p$ means that $c$ has flux $2\pi p$ around it.  Therefore, for nonzero $p$, the scalar field $\Phi$, which is charged under $c$, is twisted around the operator.  Consequently, $\Phi$ vanishes at the core of the operator.  And since the operator has an explicit factor of $\Phi^n$, the expectation value of this operator must vanish at weak coupling.  This conclusion does not apply for $p=0$.  And indeed, in Section \ref{piszero}, we will see that the behavior of the theory with $p=0$ is quite different from the behavior for nonzero $p$, and $\cal O$ can have a nonzero expectation value when $p=0$.
\end{itemize}

\subsection{The phases}\label{generalphases}

Here, we discuss the phases for nonzero $p$.  We will analyze the phases for $p=0$ in Section \ref{piszero}.  Depending on the sign of $\mu^2$ in \eqref{Higgstheory}, the theory has two phases.

\begin{itemize}
\item  For $\mu^2>0$, we can neglect $\Phi$ and we end up at low energies with the TQFT $\cal T$. 
\item  For $\mu^2<0$, $\Phi$ Higgses $c$ to a trivial field, and we end up at low energies with the TQFT ${\cal T}'$.
\end{itemize}

Importantly, the global $\U_{\cal B}$ symmetry is unbroken in these two phases.  Still, it plays an important role in the transition.  To see it, we couple the current $J_{\cal B}={\ell\over n}\mathscr{J}(\Phi)$ in \eqref{JBell} to a background field $\cal B$.  For simplicity, let it be topologically trivial. Up to using the equations of motion, this amounts to changing the kinetic term of $\Phi$ to $|D_{c+{\ell\over n}{\cal B}}\Phi|^2$.  (As we commented in footnote \ref{fractionachargeB}, the fractional gauge coupling does not affect gauge-invariant operators.)

In the $\mu^2>0$ phase, $\cal B$ plays no role.  However, in the $\mu^2<0$ phase, the Higgs fields set $c=-{\ell\over n}{\cal
B}$, up to gauge transformations.  This means that the remaining TQFT leads to the response theory ${\ell^2\over 4\pi np}{\cal B}d{\cal B}$.

Using our conventions, the response theory is expressed in terms of the Hall conductivity as $-{\sigma_H^{\cal B}\over 4\pi} {\cal B}d{\cal B}$.  (This corresponds to $J_{\cal B}={\sigma_H^{\cal B}\over 2\pi}d{\cal B}$.)  Therefore, the Hall conductivity of $\cal B$ changes across the transition\footnote{The value of $\sigma_H^{\cal B}$ suffers from two ambiguities.  First, the coupling of the background $\cal B$ to the theory can be shifted by terms that vanish on-shell.  An example is the coupling of $\cal B$ via enrichment using a one-form global symmetry  \cite{etingof2010fusion, barkeshli_symmetry_2019, Fidkowski:2016svr}.  Second, $\sigma_H^{\cal B}$ can be shifted by an even integer corresponding to adding to the Lagrangian a counterterm proportional to ${1\over 2\pi} {\cal B}d{\cal B}$.  Physically, this corresponds to stacking an SPT or, equivalently, an integer Quantum-Hall state.  (Since we discuss bosonic theories, this ambiguity is by an even integer.)  Importantly, these ambiguities do not affect the difference in the response between the two phases.\label{changeres}}
\ie\label{Deltasigma}
  \sigma_H^{\cal B}(\mu^2>0)-\sigma_H^{\cal B}(\mu^2<0)= {\ell^2\over np}\,.  
\fe
In particular, it means that even though the $\U_{\cal B}$ symmetry is unbroken, it still participates in the transition. (See \cite{Closset:2012vp} for a related discussion about comparing such Chern-Simons terms of background fields.) 

Finally, let us examine how the one-form symmetry of the theory acts in the two phases.  Recall that this symmetry is generated by all the lines $s\in \cal T$ that braid trivially with $a$.  For $\mu^2>0$, the low-energy theory is $\cal T$, and it includes these lines.  Hence, the symmetry acts faithfully.  For $\mu^2<0$ the low-energy theory is $\cal T'$.  The symmetry lines $s\in \cal T$ are mapped in $\cal T'$ to lines of the form $(s, 0)$.  The quotient in \eqref{Cprimed} does not remove any of them, nor does it identify them with other lines. Therefore, these lines act faithfully in ${\cal T}'$.  We conclude that the entire one-form symmetry of the theory acts faithfully in the low-energy theory in the two phases, and is always spontaneously broken.

\section{$p=0$}\label{piszero}

Above, we considered the case with nonzero $p$.  Here, we study the interesting special case of $p=0$. Now, ${\cal T}'$ is not a TQFT.  If the couplings to $c$ and $\Phi$ are absent, this theory is singular without a kinetic term for $\check  b$.  However, as we will see, when $c$ and $\Phi$ are present, there is no need to change our Lagrangian.

\subsection{Simplifying the theory}\label{simplifyingt}

Since $p=0$, there is no pure $\check  b$ Chern-Simons term ${\check  b}d{\check  b}$ in the Lagrangian.  This allows us to simplify it.  Following \eqref{frakbcyc}, we separate $\check  b$ into a $\U$ gauge field $b$ and ${\mathbf g}^{(2)}$.  The coupling ${n\over 2\pi} cdb$ means that $c $ can be replaced with a $\bZ_n$ gauge field ${\mathbf c}^{(1)}$
\ie
\int_{{\cal C}_1} {\mathbf c}^{(1)} = {n\over 2\pi}\int_{{\cal C}_1} c  \bmod n\,
\fe
for every 1-cycle ${{\cal C}_1}$.

Then, the kinetic term of $\Phi$ is $|D_{{2\pi \over n} {\mathbf c}^{(1)}}\Phi|^2$, and the coupling of the $\bZ_n$ gauge field ${\mathbf c}^{(1)}$ to the TQFT $\cal T$ is through
\ie\label{onetwof}
{2\pi \over n} {\mathbf c}^{(1)}\cup {\mathbf g}^{(2)}\,.
\fe
As a result, for $p=0$, we can replace $\check  b$ and $c$ with the discrete gauge fields  ${\mathbf g}^{(2)}$ and ${\mathbf c}^{(1)}$.

In this presentation of the theory,  the operator \eqref{monopoleo} is the $\bZ_n$ gauge-invariant operator
\ie\label{monopolefrakf}
{\cal O}= \Phi^{n}\,.
\fe
And as a check, it has charge $\ell$ under the $\U_{\cal B}$ current \eqref{JBell} $J_{\cal B}={\ell\over n}\mathscr{J}(\Phi)=\star{i\ell\over n}\Phi\overset\leftrightarrow{D} \bar\Phi$, which is the smallest charge for a local operator.

This conclusion can be phrased as follows.  Let us first set the ordinary $\bZ_n$ gauge field ${\mathbf c}^{(1)}$ to zero.  Then, the theory factorizes into two decoupled factors:
\begin{itemize}
    \item The first one is $\cal T$ coupled to the dynamical field ${\mathbf g}^{(2)}$, which gauges its $\bZ_n^{(1)}$ global symmetry, leading to ${{\cal T}\over \bZ_n}={{\cal T}\over (a)}$. (See Appendix \ref{sec:gauging-one-form} for a review of the gauging.)  This is a consistent quotient because $p=0$.
    \item A decoupled theory of $\Phi$.
\end{itemize}
The $\bZ_n$ gauge field ${\mathbf c}^{(1)}$ couples these two decoupled sectors.

\subsection{Local dynamics}\label{localdynpzero}

Locally, we can neglect the discrete field ${\mathbf c}^{(1)}$ and the first factor ${{\cal T}\over \bZ_n}={{\cal T}\over (a)}$, and learn that the theory includes only the second factor, which describes $\Phi$.

We recognize this second factor as an XY theory with a global $\U$ symmetry generated by
\ie\label{UfromPhi}
&W_\Phi(\theta)=\exp\left(i\theta\int_{{\cal C}_2} \mathscr{J}(\Phi)\right)\\
&W_\Phi(2\pi)=1\,.
\fe
This theory has two phases: a gapped, unbroken-symmetry phase for $\mu^2 > 0$ and a gapless, spontaneously broken symmetry phase for $\mu^2 < 0$.

If we also deform the theory by the potential term \eqref{monopolefrakf}, the global $\U$ symmetry \eqref{UfromPhi} is explicitly broken to $\bZ_n\subset\U$ generated by
\ie\label{ZnfromPhi}
&W_\Phi\left({2\pi \over n}\right)=\exp\left({2\pi i\over n}\int_{{\cal C}_2} \mathscr{J}(\Phi)\right)\\
&W_\Phi\left({2\pi\over n}\right)^n=1\,.
\fe
Then, the transition involves the spontaneous breaking of this global $\bZ_n$ symmetry.

We conclude that in this case of $p=0$,  locally, we can neglect the field ${\mathbf c}^{(1)}$ and the transition is a condensation transition \cite{Senthilunpub}. (See footnote \ref{nocondensation}.) We emphasize, though, that this conclusion is not valid globally. Examples of such transitions have been studied in various models of topological phases \cite{Fradkin:1978dv, Banks:1979fi,read_large-n_1991, Chubukov1994601,  BurnellSimonSlingerland2012}.

Interestingly, this conclusion is independent of the details of the TQFT $\cal T$ and the anyon $a$, and depends only on $n$ and on whether we broke the global $\U_{\cal B}$ symmetry by deforming the theory by an operator like $\cal O$.  Compare with Section \ref{localdynamics}.

\subsection{The zero-form symmetry }\label{zeropiszero}

The presentation of the theory at the end of Section \ref{simplifyingt}, in terms of two almost decoupled sectors, provides a new perspective on the zero-form symmetry discussed in Section \ref{zeroforms}.

As in Section \ref{simplifyingt}, let us first ignore the gauge field ${\mathbf c}^{(1)}$ that couples the two sectors.

The first sector is ${{\cal T}\over \bZ_n}={{\cal T}\over (a)}$.  And the gauging of $\bZ_n^{(1)}$ leads to a dual $\bZ_n^{(0)}$ symmetry generated by
\ie\label{ZnfromT}
W_{\cal T}=\exp\left({2\pi i\over n}\int_{{\cal C}_2} {\mathbf g}^{(2)}\right)\,.
\fe
This symmetry might not act faithfully on the TQFT ${\cal T}\over \bZ_n$.  Only a $\bZ_{\ell}^{(0)}$ quotient of it with some $\ell$ a factor of $n$ acts faithfully, i.e.,
\ie
W_{\cal T}^{\ell}=1\,.
\fe
In one extreme, as in Abelian theories, $\ell=1$, and this symmetry does not act at all.  (See Section \ref{UoneB}.) In the other extreme, as in SU(2)$_k$ with $k\in 4\bZ$, we have $\ell=n=2$, and this $\bZ_n^{(0)}$ symmetry acts faithfully by permuting some anyons.  (See Section \ref{SU2B} and note that in that case, $p=0$ is possible only when $k\in 4\bZ$.)

The second sector that describes $\Phi$ has a global $\U^{(0)}$ symmetry \eqref{UfromPhi}.  If we also add the ${\cal O}=\Phi^n$ potential, the  $\U$ symmetry \eqref{UfromPhi} is broken to $\bZ_n^{(0)}$ generated by \eqref{ZnfromPhi}.

Now, let us couple the two sectors to the same $\bZ_n$ gauge field ${\mathbf c}^{(1)}$.   This is done by replacing the kinetic term  $|\partial \Phi|^2$ by  $|D_{{2\pi \over n} {\mathbf c}^{(1)}}\Phi|^2$ and adding the coupling ${2\pi \over n} {\mathbf c}^{(1)}\cup {\mathbf g}^{(2)}$.

After the gauging, 
\ie\label{symmetryg}
W_{\cal T}=W_\Phi\left({2\pi \over n}\right)  \,.
\fe
Therefore, the unbroken symmetry is $\U_{\cal B}^{(0)}={\U^{(0)}\times \bZ_\ell^{(0)}\over \bZ_n}$ with the symmetry operators
\ie
&W_\Phi(\theta)=\exp\left(i\theta\int_{{\cal C}_2} \mathscr{J}(\Phi)\right)\\
&W_\Phi\left({2\pi \ell \over n}\right)=1\,,
\fe
i.e., the properly normalized conserved current is
\ie
J_{\cal B}={\ell\over n}\mathscr{J}(\Phi)\,.
\fe
Its $\bZ_\ell^{\cal B}\subset \U_{\cal B}^{(0)}$ symmetry, which is generated by $W_{\cal T}$ of \eqref{ZnfromT} acts only on line operators.

If we also add the ${\cal O}=\Phi^n$ potential, this $\U_{\cal B}^{(0)}$ symmetry is explicitly broken to $\bZ_\ell^{\cal B}$ of \eqref{ZnfromT} that acts only on line operators. Given that $\Phi$ creates the anyon $a$, adding $\cal O$ to the Lagrangian means that anyon $a$ satisfies $a^n={\bf 1}$ in the full transition theory and not only in the TQFT $\cal T$.

\subsection{The phases}\label{piszerop}

Let us turn to the phases of the theory.  As for nonzero $p$, the low-energy limit of the positive $\mu^2$ phase is the TQFT $\cal T$.

The situation for negative $\mu^2$ is more interesting.   We expand around $\Phi=\rho \exp(i\alpha)$ with nonzero $\rho$.  Since $\Phi$ couples to a discrete gauge field ${\mathbf c}^{(1)}$, $\alpha$ is  massless; i.e., the theory is gapless.  

We recognize $\alpha$ as the Goldstone boson of the spontaneously broken $\U_{\cal B}$ symmetry.  Near the end of Section \ref{zeroforms}, we argued that for nonzero $p$, at weak coupling, the operator $\cal O$ cannot have a nonzero expectation value.  As is clear from \eqref{monopolefrakf}, for $p=0$ and negative $\mu^2$, $\langle {\cal O}\rangle\ne 0$, thus breaking the global $\U_{\cal B}$ symmetry.  And $\alpha$ is its Goldstone mode.

The massless field $\alpha$ couples to the modes of the TQFT $\cal T$ through \eqref{onetwof}. More explicitly, $\oint {\mathbf c}^{(1)}={n\over 2\pi}\oint d\alpha$, and therefore, the winding modes (vortices) of $\alpha$ are charged under ${\mathbf g}^{(2)}$.\footnote{As a field, $\alpha \sim \alpha+2\pi$.  But since the $\bZ_n$ gauge symmetry relates $\alpha$ to $\alpha +{2\pi \over n}$, the phrase ``winding'' here corresponds to $\alpha$ winding to  $\alpha +{2\pi \over n}$.\label{windingnor}}  Hence, depending on the winding (vorticity) modulo $n$, the vortex carries different anyons in $\cal T$.

Let us elaborate on this winding charge in the low-energy theory of the gapless phase.  The exact $\bZ_n^{(1)}$ symmetry generated by the anyon $a$, or equivalently by the line $e^{i\oint c}=e^{{2\pi i\over n}\oint {\mathbf c}^{(1)}}$ acts at low energies as $e^{i\oint d\alpha}$.  It is enlarged to an emergent $\U^{(1)}$ symmetry with winding (vorticity) charge ${n\over 2\pi}\oint d\alpha$. (See footnote \ref{windingnor}.)  Such a winding symmetry in a gapless $2+1$d phase should be viewed as unbroken \cite{Gaiotto:2014kfa}.

In Section \ref{SU2ellistwo}, we will demonstrate this coupling of $\alpha$ to $\cal T$ in a special case with $\ell=2$.

Next, as above, we consider deforming the model by ${\cal O}=\Phi^n$ of \eqref{monopolefrakf}, thus explicitly breaking the global $\U_{\cal B}$ symmetry to $\bZ_\ell^{\cal B}$.

For positive $\mu^2$, the deformation by ${\cal O}=\Phi^n$ is not important. But for negative $\mu^2$, it gives $\alpha$ a mass, gaps the model, and leads to ${\cal T}\over \bZ_n$.  We conclude that this transition theory implements gauging by relating $\cal T$ to ${\cal T}\over \bZ_n$.

\subsection{Comparing the transition with gauging}

The transition theory deformed by $\cal O$ implements the gauging ${\cal T}\to {{\cal T}\over \bZ_n}$.  Let us compare them in more detail.

\subsubsection{The zero-form symmetry}

After the deformation by $\cal O$, the $\U_{\cal B}$ zero-form symmetry is explicitly broken to $\bZ_\ell^{\cal B}$, which acts only on lines.  It is not spontaneously broken.

For positive $\mu^2$, this symmetry does not act on the low-energy modes of $\cal T$.  In the negative $\mu^2$ phase,  $\bZ_\ell^{\cal B}$ acts faithfully on the low-energy modes of ${\cal T}\over \bZ_n$.  

In the gauging picture, $\bZ_\ell^{\cal B}$ is the faithfully acting subgroup of the dual $\bZ_n^{(0)}$ symmetry.  Now we see that in the transition field theory, $\bZ_\ell^{\cal B}\subset \bZ_n^{(0)}$ is an exact symmetry of the full transition theory.

\subsubsection{The one-form symmetry}

Next, let us discuss the various exact one-form symmetries.  Since $p=0$, the anyon $a$ braids trivially with itself.  Therefore, the full $\bZ_n^{(1)}$ symmetry generated by $a$ is an exact symmetry of the transition theory.  Let us see how it acts in the two phases.

For positive $\mu^2$, this $\bZ_n^{(1)}$ symmetry is spontaneously broken, as it acts on the low-energy modes of $\cal T$.  However, for negative $\mu^2$, $\bZ_n^{(1)}$ is unbroken.\footnote{We mentioned above that before the deformation by $\cal O$, the low-energy theory in the negative $\mu^2$ phase had an emergent $\U^{(1)}$ winding symmetry generated by  ${n\over 2\pi}\oint d\alpha$. And this symmetry is not spontaneously broken.  The deformation by $\cal O$ explicitly breaks this emergent $\U^{(1)}$  to the exact $\bZ_n^{(1)}$ symmetry generated by $e^{i\oint c}$. (See footnote \ref{windingnor}.)  This symmetry is also not spontaneously broken in this phase.}  Indeed, it does not act faithfully on the low-energy modes, which are described by ${\cal T}\over \bZ_n$.  In the gauging picture of this phase, these symmetry lines do not act because they are identified with the trivial lines.  In the transition field theory, they do not act at low energy because the corresponding symmetry is unbroken.  Instead, they act on massive states corresponding to the winding of $\alpha$.

Let us consider the remaining one-form symmetry of the problem.  The original TQFT $\cal T$ has a certain one-form symmetry ${\cal G}^{(1)}$.  The $\bZ_n^{(1)}$ generated by $a$ is a subgroup of ${\cal G}^{(1)}$.  Since $p=0$, this subgroup is anomaly-free.  However, it has nonzero 't Hooft anomalies with other elements in ${\cal G}^{(1)}$.  When we gauge $\bZ_n^{(1)}$ to find ${\cal T}\over \bZ_n$, that 't Hooft anomaly turns into an ABJ anomaly, which explicitly breaks the original ${\cal G}^{(1)}$ to the subgroup that braids trivially with $\bZ_n^{(1)}$.  And in addition, we have a quotient by $\bZ_n^{(1)}$.

In the transition field theory, there is no ABJ anomaly.  Here, this explicit breaking of ${\cal G}^{(1)}$ to the subgroup that braids trivially with $\bZ_n^{(1)}$ is obtained through the coupling to $\Phi$. And also, as we said above, there is no quotient by $\bZ_n^{(1)}$.  Instead, $\bZ_n^{(1)}$ acts only at high energies.

Importantly, the coupling to $\Phi$ breaks a (one-form) symmetry explicitly rather than spontaneously.  For positive $\mu^2$, this explicit breaking is not important at low energies and the full ${\cal G}^{(1)}$ is an emergent symmetry.  However, for negative $\mu^2$, this explicit symmetry breaking is important at low energies. This situation is very similar to the BKT transition.  There, the UV lattice theory lacks the winding/vorticity symmetry.  In one phase, this symmetry is emergent.  And in the other phase, it is explicitly broken.  This analogy is another reason the transition should be viewed as a proliferation transition, rather than a condensation transition. (See footnote \ref{nocondensation}.)

\subsection{Summary}

In conclusion, for $p=0$, our Lagrangian with the added operator ${\cal O}$ leads to a phase transition from a phase with $\cal T$ to a phase with ${{\cal T}\over \bZ_n} ={{\cal T}\over (a)}$.  It implements the gauging of the one-form symmetry.

Intuitively, this construction can be viewed as follows.  We start with the TQFT $\cal T$ and gauge its anomaly-free $\bZ_n^{(1)}$ symmetry generated by the anyon $a$, using the gauge field  ${\mathbf g}^{(2)}$.  (It is anomaly-free because $p=0$.) This leads to a dual $\bZ_n^{(0)}$ symmetry.  We gauge it using the $\bZ_n$ gauge field ${\mathbf c}^{(1)}$ to find an alternate description of the original TQFT $\cal T$.  Then, we introduce $\Phi$ that couples to ${\mathbf c}^{(1)}$ and add a $\bZ_n$ gauge-invariant $\Phi^n$ potential. We end up with a Higgs transition for this $\bZ_n$ gauge theory. Such a construction for a Higgs transition that implements gauging of an anomaly-free $\bZ_n^{(1)}$ symmetry has been suggested by various people~\cite{receipe}.

\section{A dual description}\label{dual theory}

In preparation for dualizing our theory, we start by reviewing the standard particle/vortex duality.  We consider a complex scalar field $\chi$ with a global $\U_B$ symmetry and study its Lagrangian
\ie
|D_B\chi|^2-\mu^2|\chi|^2 +\cdots\,,
\fe
where $B$ is a background $\U_B$ gauge field.   (See footnotes \ref{absuse} and \ref{convensionsf}.) This theory is dual to a theory of a complex scalar field $\tilde \chi$ and a dynamical $\U$ gauge field $\tilde b$ with the Lagrangian
\ie
|D_{\tilde b}\tilde\chi|^2-\tilde\mu^2|\tilde \chi|^2 +|d\tilde b|^2- {1\over 2\pi} B d\tilde b+\cdots\qquad, \qquad \tilde \mu^2=-\mu^2\,.
\fe
Importantly, we need to include the kinetic term for the dynamical gauge field $\tilde b$, because otherwise, this theory is singular.  The monopole operator ${\cal M}_{\tilde b}$ of $\tilde b$ has charge $1$ under $\U_B$ (see footnote \ref{convensionsf}), and hence, we identify $ {\cal M}_{\tilde b} =  \chi$, where we continue to suppress unimportant coefficients.

Let us use this particle/vortex duality in our theory \eqref{Higgstheory} with $\Phi=\chi$, $c=B$, and $\tilde \Phi=\tilde \chi $ to find its dual version
\ie\label{}
{\cal L}_{{\cal T}'}   +{1\over 2\pi } {\frak b} dc+|D_{\tilde b}\tilde\Phi|^2 - {1\over 2\pi} c d\tilde b +|d\tilde b|^2-\tilde\mu^2|\tilde \Phi|^2 + \cdots\,.
\fe
Integrating out $c$, we learn that (up to a gauge transformation), $\tilde b ={\frak b}$, and the dual theory is
\ie\label{dualLag}
{\cal L}_{{\cal T}'}   +|D_{ {\frak b}}\tilde\Phi|^2  +|d{\frak b}|^2-\tilde\mu^2|\tilde \Phi|^2 + \cdots\,.
\fe

Using these variables, the $\U_{\cal B}$ current \eqref{JBell} can be written as
\ie\label{}
&J_{\cal B}={\ell\over n}\mathscr{J}(\Phi)= {\ell\over 2\pi n}  d{\frak b}=-{\ell \over p} \mathscr{J}(\tilde \Phi)\\
&\mathscr{J}(\tilde \Phi)= \star i\tilde\Phi\overset\leftrightarrow{D} \bar{\tilde\Phi}\,,
\fe
where we used the equations of motion and the conventions in footnote \ref{convensionsf}. For $p=0$, the current $J_{\cal B}$ exists, but it cannot be expressed only in terms of $\tilde \Phi$.

As a check, the gauge-invariant operator \eqref{monopoleo} ${\cal O}= \Phi^{n}{\cal M}_{\frak b}^n{\cal M}_c^{p}$ is expressed as
\ie\label{monopoledual}
{\cal O}= \left({\bar{\tilde\Phi}}\right)^{p}{\cal M}_{\frak b}^n\,,
\fe
and its $\U_{\cal B}$ charge is $\ell$.

The analysis of the phases using this description yields the same conclusions as above.
\begin{itemize}
\item For $\mu^2=-\tilde \mu^2>0$, $\tilde \Phi$ Higgses ${\frak b}$.  This means that the $\U_{-np}$ factor in the numerator of \eqref{Cprimed} and the  $\bZ_n$ quotient are absent.  (See the comments following \eqref{cocyclno}.) As a result, we end up at low energies with the TQFT $\cal T$. 
\item For $\mu^2=-\tilde \mu^2<0$, $\tilde \Phi$ is massive.  At low energies, we can neglect it and find the TQFT ${\cal T}'$ with a kinetic term for ${\frak b}$.  For nonzero $p$, this kinetic term can be neglected, and we find a gapped phase with ${\cal T}'$.  For $p=0$, the kinetic term is important, and we find a gapless mode.  Then, deforming the theory by ${\cal O}$ lifts that gapless mode.
\end{itemize}

This dual picture gives us more information about the transition.  The original discussion of the transition was motivated by the fact that the anyon $a\in {\cal T}$, created by the field $\Phi$, becomes light as we approach the transition.  Then it proliferates, leading to the transition.  Here we learn that the light mode after the transition is created by the field $\tilde \Phi$, which corresponds to the anyon $\tilde a=(\mathbf{1}, n)\in{\cal T}'$, which is represented by $\exp\left(i\int {\frak b}\right)$.  Therefore, the dual theory gives us a transition that implements the inverse map in Section \ref{inversemap}.

Finally, as in Section \ref{localdynamics}, the local dynamics is independent of ${\cal T}$ and is given by
\ie
-{p\over 4\pi n}{\frak b} d {\frak b} +|D_{ {\frak b}}\tilde\Phi|^2  +|d{\frak b}|^2-\tilde\mu^2|\tilde \Phi|^2 + \cdots\
\fe
where ${\frak b}$ is an $\bR$ gauge field.  Up to renaming the fields, this is the same as \eqref{dualtheory} with ${n\over p}\to -{p\over n}$. Clearly, this is consistent with the inverse map in Section \ref{inversemap}.

\section{Coupling the theory to a background $\U_A$ gauge field}\label{enrichmentsec}

In this section, we consider starting with a UV microscopic theory with a global $\U_A$ symmetry and examine its action on our transition theory.  See Figure \ref{threescales}.

\subsection{Enriching $\cal T$ by $\U_A$}\label{Enrichment}

Let us start with the positive $\mu^2$ phase with $\cal T$.  In this phase, the $\U_A$ symmetry can enrich the topological theory \cite{etingof2010fusion,barkeshli_symmetry_2019}.\footnote{Here, enrichment means that the $\U_A$ zero-form symmetry does not act faithfully in the theory.  Nevertheless, it can still act via a one-form symmetry. In TQFTs, this phenomenon is also known as ``symmetry fractionalization.'' In the condensed matter literature, it is common to combine the zero-form symmetry that acts faithfully (e.g., by permuting lines) and the enrichment, and to refer to the TQFT as ``symmetry-enriched topological order.''}   This happens by selecting an Abelian anyon $v\in {\cal T}$ to act as the vison.  This means that the $\U_A$ charge of any anyon $s\in {\cal T}$ is determined by its braiding with $v$, $\exp\left(2\pi i\langle v,s\rangle\right)$ as
\ie\label{QAofs}
Q_A(s)=\langle v,s\rangle\bmod 1\,.
\fe
Importantly, this charge is meaningful only modulo one.

The coupling of the theory to a background gauge field $A$ for $\U_A$ allows us to define the Hall conductivity of $A$, $\sigma_H^A$, through the response theory $-{\sigma_H^{A}\over 4\pi} AdA$. It is related to the spin of the vison as
\ie\label{sigmaAv}
\sigma_H^A(\mu^2>0)=2h(v)\bmod 2\,.
\fe
As in footnote \ref{changeres}, the modulo 2 ambiguity arises from a possible added counterterm ${2\over 4\pi} AdA$ (equivalently, stacking a bosonic $\U_A$ SPT state).

Using \eqref{QAofs}, the $\U_A$ charge of our special anyon $a\in {\cal T}$ is
\ie\label{QAofa}
Q_A(a)=\langle v,a\rangle\bmod 1\,.
\fe
Since $a$ generates a $\bZ_n^{(1)}$ symmetry, then, \eqref{QAofa} means that the vison $v$ transforms under this symmetry as
\ie
\exp\left(2\pi i Q_A(a)\right)=\exp\left({2\pi i n_v\over n}\right)\,.
\fe
(See Section \ref{calT}.)  We learn that the $\U_A$ charge of $a$ is related to the $\bZ_n^{(1)}$ representation of $v$.

\subsection{Coupling the transition theory to $\U_A$}

The UV theory has a conserved $\U_A$ current $J_A$.  In the low-energy phase with $\cal T$, this current vanishes.  (Recall that operators are always defined modulo the equations of motion.  So when we say that $J_A$ vanishes, we mean that it vanishes on-shell.)  This is the statement that $\U_A$ couples to $\cal T$ only through enrichment via the one-form symmetry generated by the vison $v$.  

Now, we would like to find how the UV current $J_A$ acts in the transition theory.  Since this theory has only one conserved current, $J_{\cal B}$, on-shell $J_A$ should be proportional to $J_{\cal B}$.  To find the proportionality factor, we use the fact that in the phase with $\cal T$, $Q_A(a)={n_v\over n}\bmod 1$. This determines
\ie\label{JAJbo}
J_A= {n_v\over \ell}J_{\cal B}={n_v\over n } \mathscr{J}(\Phi)= {n_v\over 2\pi n}  d{\frak b}\,.
\fe

More precisely, up to this point, only $n_v\bmod n$ was defined and used. The remaining freedom to shift $n_v$ by ${\cal Q}n$ with an integer $ {\cal Q} $ corresponds to changing the coupling of the background gauge field $A$ to $\Phi$, i.e., changing the covariant derivative of $\Phi$ to $D_{c+{\cal Q}A}\Phi$. Here, we absorb the freedom in this coupling in the value of $n_v$ as an integer.\footnote{If, as in Figure \ref{threescales}, we start with a given UV theory with a specific $\U_A$ symmetry, the current $J_A$ is fixed by the UV theory, and there is no freedom to change the coupling of $A$ to $\Phi$.  When we refer to freedom to change it or say that it is ambiguous, we discuss it from the IR perspective without the constraints from the UV.\label{UVcalQ}}

This leads to the conclusion that the $\U_A$ charge of $\cal O$ is
\ie\label{calOAcharge}
Q_A({\cal O})=n_v\,.
\fe
As a check, this charge is quantized.

As a more subtle check, \eqref{JAJbo} means that the $\U_A$ charge of ${\cal O}'$ of \eqref{monopoleopp} is
\ie\label{QofOp}
Q_A({\cal O}')={n_v\over \ell}\,.
\fe
$\cal O'$ corresponds to a state on a Riemann surface ${\cal C}_2$ with the anyon  $a^{n\over \ell}$ coming out of it.  The braiding of $v$ with this line should be trivial.  Therefore, $n_v\over \ell$ should be an integer.  This guarantees that $Q_A({\cal O}')={n_v\over \ell} \in \bZ$.

When $Q_A(a)$ is an integer, $n_v=0\bmod n$, and we can set $\cal Q$ such that $n_v=0$.  Equivalently, in this case, the vison $v$ braids trivially with $a$ and it generates a global one-form symmetry of the whole transition theory.  Setting $n_v=0$ means that $J_A$ vanishes on-shell and $\U_A$ couples to the transition theory only through enrichment via this one-form symmetry.  In particular, in that case, $Q_A({\cal O})=0$.

However, this is not the case when $n_v\ne 0$.  This can happen either because $Q_A(a)=0\bmod 1$, but we did not set $\cal Q$ such that $n_v=0$, or because $Q_A(a)$ is fractional.  Either way, when $n_v\ne 0$, $J_A$ does not vanish on-shell and $\cal O$ necessarily carries nonzero $\U_A$ charge.

To conclude, in the TQFT, $\U_A$ does not act faithfully.  It acts only as enrichment, and the current $J_A$ vanishes.  Related to that, the charges of the anyons are fractions that are well-defined only modulo 1 \eqref{QAofs}.  This is to be contrasted with the situation for nonzero $n_v$ in the transition theory.  Here, $\U_A$ acts faithfully, the current $J_A$ is nonzero \eqref{JAJbo}, and the charges $Q_A$ are well-defined integers, e.g., \eqref{calOAcharge}.  In addition, line operators in this phase can have fractional charges.

\subsection{Enriching $\cal T'$  by $\U_A$}

Next, we consider the negative $\mu^2$ phase, and we limit ourselves to nonzero $p$. (We will return to $p=0$ below.)  Here, the low-energy theory is described by \eqref{Cprimed}
\ie\label{}
{\cal T}'={{\cal T}\boxtimes \U_{-pn}\over \bZ_n} = {{\cal T}\boxtimes \U_{-pn}\over (a,p)}\,.
\fe

The background gauge field $A$ can couple to $\cal T'$ only via enrichment.  So we should identify the vison $v'\in {\cal T}'$.  The form of the current in the transition theory \eqref{JAJbo}
$J_A=  {n_v\over 2\pi n}  d{\frak b}={n_v\over 2\pi}  d{\check b}$ means that $v'$ should act on anyons of the form $(1,w_b)\in {\cal T}'$ as  $\exp\left(in_v\int \check b\right)$. (See footnote \ref{differentbs}.)  Combining this with the fact that it should act on anyons of the form $(s,0)\in {\cal T}'$ as $v$, leads us to identify 
\ie\label{vpv}
v'=(v,n_v)\in {\cal T}'\,.
\fe
As a check, this anyon braids trivially with $(a,p)$ and therefore it is indeed in $\cal T'$.

Interestingly, $n_v$ and $n_v+n$ do not lead to the same vison $v'$.  Therefore, even though they correspond to the same charge assignments in $\cal T$, they lead to different charge assignments in $\cal T'$. In more detail, the coupling of $A$ to the transition theory depends on the integer $n_v$.  The coupling of $A$ to $\cal T$ depends only on $n_v\bmod n$.    And the coupling to $\cal T'$ depends on $n_v\bmod np$.  This is less information than the coupling in the transition theory, but it can be more information than in the coupling to $\cal T$.

As in \eqref{sigmaAv}, now we have
\ie\label{}
\sigma_H^A(\mu^2<0)=2h(v')\bmod 2\,.
\fe
Given \eqref{vpv}, we learn that
\ie
\sigma_H^A(\mu^2<0)=\left(\sigma_H^A(\mu^2>0)-{n_v^2\over pn} \right)\bmod 2\,.
\fe
In fact, we have a stronger result.  The modulo 2 ambiguity in $\sigma_H^A$ arises from adding a local counterterm ${2\over 4\pi} AdA$.  This freedom is the same on both sides of the transition.  Therefore, we learn that 
\ie\label{changesigma}
\sigma_H^A(\mu^2<0)=\sigma_H^A(\mu^2>0)-{n_v^2\over pn} \,.
\fe
Alternatively, we can derive this result using the similar relation for $\sigma_H^{\cal B}$ in \eqref{Deltasigma} and the expression for the current $J_A={n_v\over \ell}J_{\cal B}$ in \eqref{JAJbo}.

\subsection{Coupling the $p=0$ theory to $\U_A$}

The theories with $p=0$ are particularly interesting because they can implement gauging a one-form symmetry as a transition.  Here, we discuss how they are affected by the coupling to $\U_A$.

If $n_v=0$, $\U_A$ acts as pure enrichment of the full transition theory, $J_A=0$, and $Q_A({\cal O})=0$.  Therefore, the coupling to the background field $A$ does not affect our previous discussion.

Nonzero $n_v$ can arise either when $Q_A(v)\ne 0\bmod 1$, or when $Q_A(v)= 0\bmod 1$, but the coupling of $\Phi$ to $A$ leads to nonzero $n_v$.  In these two cases, $\U_A$ acts faithfully on the transition theory, $J_A\ne 0$, and $Q_A({\cal O})=n_v\ne 0$.  

Recall that the $p=0$ theory has two versions, with or without adding $\cal O$ to the Lagrangian.  The fact that $Q_A({\cal O})=n_v\ne 0$ is quite important in these two versions.
\begin{itemize}
\item If we do not add $\cal O$ to the Lagrangian, the negative $\mu^2$ phase is gapless, and the global $\U_{\cal B}$ symmetry is spontaneously broken. For nonzero $n_v$, this also means that $\U_A$ is spontaneously broken.
\item If we do add $\cal O$ to the Lagrangian, the global $\U_{\cal B}$ symmetry is explicitly broken. And that means that the theory does not respect the UV $\U_A$ symmetry.
\end{itemize}

The version of the $p=0$ transition theory with an added $\cal O$ term in the Lagrangian implements gauging the one-form symmetry generated by $a$.  Now we see that if $n_v\ne 0$, this theory explicitly breaks $\U_A$ and therefore, it is incompatible with enriching $\cal T$.

As we said in footnote \ref{UVcalQ}, starting with a given UV theory, the value of $\cal Q$ and therefore also the value of $n_v$ are determined.  Then, if $n_v\ne 0$, the UV $\U_A$ symmetry prevents the presence of $\cal O$ in the Lagrangian, and the $p=0$ theory does not lead to gauging.  It leads to a gapless phase with spontaneously broken $\U_A$.

If, however, we do not insist on a specific UV theory, we can still adjust the system so that it has a transition description of gauging its $\bZ_n^{(1)}$ and be coupled to $\U_A$.

One way this can happen is when $n_v=0\bmod n$.  In this case, we can change the coupling of $A$ to $\Phi$ by changing $\cal Q$ and setting $n_v=0$. Then, as above, we can add $\cal O$ to the Lagrangian, break $\U_{\cal B}$, and find a gapped phase with ${\cal T}\over \bZ_n$, while preserving $\U_A$.  In this case, $\sigma_H^A$ does not change across the transition.

When $n_v\ne 0\bmod n$, a more drastic modification is needed.  As is common in studies of TQFT, we should extend $\U_A$.  (See Appendix \ref{enrichementapp}.) This extension corresponds to changing the charge normalization of $\U_A$.  Specifically, instead of taking the vison $v$, we take the vison $\hat v=v^{n\over \gcd(n,n_v)}$.  Since the braiding $\langle \hat v, a\rangle ={n_v\over \gcd(n,n_v)}\bmod 1=0\bmod 1$ is trivial, $n_{\hat v}=0\bmod n$.  With appropriate $\cal Q$, $n_{\hat v}=0$.  Then, $\U_{\hat A}$ acts as pure enrichment of the whole transition theory.  The only difference relative to the case with $\U_A$ is that the charges of the anyons are larger
\ie
Q_{\hat A}(s) = {n\over \gcd(n,n_v)}Q_A(s)\bmod 1\,.
\fe
Consequently, the fluxes of the background gauge field $A$ are restricted:
\ie
\int_{{\cal C}_2}{dA\over 2\pi}={n\over \gcd(n,n_v)}\int_{{\cal C}_2}{d\hat A\over 2\pi}\in {n\over \gcd(n,n_v)}\bZ\,.
\fe
Our discussion here shows how this extension can be carried out in the full transition theory, and not just in the TQFT.

\section{Examples}\label{twoexamples}

In the previous sections, we gave a general treatment of our problem.  Here, we will illustrate it with two simple examples.  In Section \ref{Abelianexamples}, we will discuss Abelian theories, and in Section \ref{TisSU2}, we will discuss SU(2) gauge theories.

One advantage of the concrete examples is that here, we can use standard gauge fields to present quotients like \eqref{CCCrel} without the need for twisted gauge fields like $\check  b$ above.

\subsection{$\cal T$ is an Abelian TQFT}\label{Abelianexamples}

\subsubsection{$\cal T$ and ${\cal T}'$}\label{TTpAbelian}

Here, we study a general Abelian TQFT $\cal T$ \cite{Belov:2005ze, Stirling:2008bq,Kapustin:2010hk}.  It can be described by a $\U^r$ gauge theory with gauge fields $b^i$ and the Lagrangian
\ie\label{TAbeLag}
&{\cal L}_{\cal T}={1\over 4\pi} \sum_{i,j=1}^r b^i K_{ij}db^j\\
&K_{ii}\in 2\bZ \qquad, \qquad K_{ij}=K_{ji}\in \bZ \qquad,\qquad \det K\ne 0\,,
\fe
where we included the restriction that the theory is bosonic.

An example of the fact mentioned in the introduction, that the same $\cal T$ can be described by many different gauge groups, is the addition of two more $\U$ gauge fields with the change
\ie\label{adddecou}
K\to \begin{pmatrix}
    K&0&0\\
    0&0&1\\
    0&1&0
\end{pmatrix}\,.
\fe

The anyons in the theory are labeled by a vector $w\in \bZ^r$ and the corresponding line operator is the Wilson line $\exp\left(i\sum_i w_i\int b^i\right)$.  The fact that the same anyon can correspond to different representations of the gauge group is demonstrated by the identification
\ie\label{wident}
w\sim w+K{\bf l}\qquad, \qquad {\bf l}\in \bZ^r\,,
\fe
where we used vector notation.
The spin of the anyon is
\ie
h(w)={1\over 2}w^{\mathrm T}K^{-1}w\bmod 1\,,
\fe
where we continue to use the vector notation.  It is easy to check using \eqref{TAbeLag} that the identification \eqref{wident} is compatible with $h(w)$.

The special anyon we focus on, $a$, is labeled by a vector $q\in \bZ^r$.  For the moment, we are going to ignore the identification \eqref{wident} for $q$.  Below, we will discuss the dependence on ${\bf l}$.

Given $q$, the order of the anyon $n$ is the smallest positive integer such that
\ie\label{udefe}
u=n K^{-1} q\in \bZ^r\,,
\fe
and we define
\ie\label{pualA}
p=q^{\mathrm T} u=n q^{\mathrm T} K^{-1} q\,.
\fe
As above, the anomaly is $p\bmod 2n$ \cite{Hsin:2018vcg}, but we view $p$ as an integer.  Related to that, under $q\to q+K{\bf l}$, we have $u\to u+n {\bf l}$ and $p\to p+2 n q^{\mathrm T}{\bf l}+n {\bf l}^\mathrm{T}K{\bf l}$.  In particular, $q$ and $q+K{\bf l}$ with $2 q^{\mathrm T}{\bf l}+ {\bf l}^\mathrm{T}K{\bf l}=0$ have the same $p$.\footnote{The addition of two more $\U$ gauge fields as in \eqref{adddecou} does not affect $\cal T$ and therefore, it does not affect $p\bmod 2n$.  However, in some cases, not all values of $p$ with the same $p\bmod 2n$ can be achieved without adding these fields.  Below, we will assume that if needed, such fields were added.  See \cite{CStoappear} for a detailed discussion.}

Now, given this $\cal T$ and the anyon $a$ labeled by $q$, we follow the general discussion above.

First, we consider the case with nonzero $p$.  To find a Lagrangian for ${\cal T}'$ \eqref{Cprimed}, we add to it
\ie
{\cal L}_{\U_{-np}}=-{np\over 4\pi} bdb\,,
\fe
and then perform the quotient by $(a,p)$.

The quotient in \eqref{Cprimed} is achieved by replacing $b^i$ and $b$ by twisted gauge fields $\check {b}^i$ and $\check {b}$, whose fluxes can be fractional
\ie\label{Abelianfrac}
        &{1\over 2\pi}\int_{{\cal C}_2} d \check {b}^i+ {u^i\over 2\pi}\int_{{\cal C}_2} d \check {b}\in \bZ\\
        &{1\over 2\pi}\int_{{\cal C}_2} d \check {b}\in {1\over n}\bZ\,.
\fe
As a check, the anyon $(a,p)$, which is described in the numerator, ${\cal T}\boxtimes \U_{-np}$, by the Wilson line $\exp\left(i\int \left(\sum_{i} q_i {b}^i +  p {b}\right)\right) $ should be transparent in the quotient.  Indeed, it generates a ``Dirac string'' with fractional fluxes satisfying \eqref{Abelianfrac}.

In terms of these twisted gauge fields,
\ie\label{frakbLagA}
{\cal L}_{{\cal T}'}={1\over 4\pi} \sum_{i,j=1}^r \check {b}^i K_{ij}d\check {b}^j-{np\over 4\pi} \check {b}d\check {b}\,.
\fe
As we mentioned in the introduction to this section, here, we can avoid using the twisted gauge fields $\check {b} ^i$ and $ \check {b}$.  This Lagrangian for ${\cal T}'$ can be expressed using standardly normalized gauge fields ${\frak b}^i$ and ${\frak b}$ as
\ie\label{AbeKprime}
&{\cal L}_{{\cal T}'}={1\over 4\pi} \sum_{i,j=1}^r {\frak b}^i K_{ij}d{\frak b}^j-{1\over 2\pi} \sum_{i=1}^r q_i {\frak b}^id{\frak b}\\
    &{\frak b}^i= \check {b}^i+u^i \check {b}\\
    &{\frak b}=n \check {b}\,.
\fe
Recall our notation in footnote \ref{differentbs}.

  We recognize $\cal T'$ as a $\U^{r+1}$ Chern-Simons theory with a K-matrix
\ie\label{Krp}
&K'=\begin{pmatrix}
        0&-q^\mathrm{T}\\-q&K
    \end{pmatrix}\\
    &\det  K'=-q^{\mathrm T} K^{-1}q \det K =-{p\over n}\det K\,.
\fe
As a check, the determinant vanishes when $p=0$.

As another check, the Lagrangian \eqref{AbeKprime} is invariant under $q_i\to q_i+ \sum_j K_{ij}{\bf l}^j$ combined with ${\frak b}^i\to{\frak b}^i+{\bf l}^i{\frak b}$, provided $2 q^{\mathrm T}{\bf l}+ {\bf l}^\mathrm{T}K{\bf l}=0$.  This shows that the only information in $q$ beyond $q\sim q+ K{\bf l}$ is included in the value of $p$.

Finally, the Lagrangian \eqref{AbeKprime} demonstrates one of the points in Section \ref{TTprel}.  The first term in \eqref{AbeKprime} depends on the $\U^r$ gauge fields ${\frak b}^i$.  It describes the original TQFT $\cal T$.  If the $\U$ gauge field ${\frak b}$ had been a classical gauge field, denoted $-X$ in Section \ref{TTprel}, this term would have described an enrichment using the anyon $a$, as its vison.  Instead, the theory $\cal T'$ is found by coupling $\cal T$ to such a background field and then making it dynamical.

\subsubsection{The non-topological transition field theory}\label{AbelianUVtheory}

Next, as in \eqref{Higgstheory}, we add the $\U$ gauge field $c$  and the scalar $\Phi$ and write
\ie\label{btildeLagp}
{1\over 4\pi} \sum_{i,j=1}^r {\frak b}^i K_{ij}d{\frak b}^j-{1\over 2\pi} \sum_{i=1}^r q_i {\frak b}^id{\frak b}+{1\over 2\pi} {\frak b}dc+|D_{c}\Phi|^2 -\mu^2|\Phi|^2 + \cdots\,.
\fe
For $p=0$, the TQFT ${\cal T}'$ based on the Lagrangian \eqref{AbeKprime} is singular unless we add a kinetic term for ${\frak b}$.  However, as we said above, there is no need to add this term explicitly in the theory \eqref{btildeLagp}, even when $p$ is zero.

 As for the TQFT $\cal T'$
\eqref{AbeKprime}, the theory based on \eqref{btildeLagp} is independent of the shift $q\to q+K{\bf l}$, as long as this shift preserves $p$.

Since ${\frak b}$ is a standard $\U$ gauge field and there is no ${\frak b} d{\frak b}$ term in the Lagrangian, we can integrate it out to find (up to a gauge transformation)
\ie\label{cinterbi}
c=\sum_i q_i {\frak b}^i\,.
\fe
We end up with $\U^r$ gauge fields ${\frak b}^i$ coupled to $\Phi$ with charges $q_i$.  The Lagrangian is
\ie\label{btildeLag}
{1\over 4\pi} \sum_{i,j=1}^r {\frak b}^i K_{ij}d{\frak b}^j+|D_{\sum_i q_i{\frak b}^i}\Phi|^2 -\mu^2|\Phi|^2 + \cdots\,.
\fe

This fits our general picture. The first term represents our starting TQFT $\cal T$, and the other terms show that $\Phi$ is a scalar field corresponding to the anyon labeled by $q$.  An earlier version of such a transition theory was written in \cite{WenWu1993}.

\subsubsection{$\U_{\cal B}$}\label{UoneB}

Let us determine the conserved current $J_{\cal B}$.  To do that, it is convenient to return tentatively to the presentation \eqref{btildeLagp} before integrating out ${\frak b}$ and $c$.

Following Section \ref{zeroforms}, the theory has a zero-form symmetry $\U_{\cal B}$ with the current \eqref{JBell}
\ie\label{currentAbel}
J_{\cal B}={\ell\over n}\mathscr{J}(\Phi)= {\ell\over 2\pi n}  d{\frak b}\,,
\fe
where $\ell|L$ is determined by the on-shell flux quantization \eqref{restfrombg}
\ie\label{fluxab}
{1\over 2\pi}\int_{{\cal C}_2}d{\frak b}\in {n\over \ell}\bZ\qquad, \qquad \ell|L\,.
\fe
In order to find the quantization of these on-shell fluxes, we use the Lagrangian \eqref{btildeLagp}. Here, the equations of motion of ${\frak b}^i$, set $nd{\frak b}^i=u^i d{\frak b}$, where we used \eqref{udefe}.  The equation \eqref{udefe} also implies that
\ie\label{gcdnu}
\gcd(n,u^1,u^2,\cdots,u^r)=1\,.
\fe
Combining this with the fact that the fluxes of ${\frak b}^i$ have standard quantization, the on-shell fluxes of ${\frak b}$ should be multiples of $n$, corresponding to $\ell=1$  in \eqref{fluxab}.  As a result, the current \eqref{currentAbel} is
\ie\label{JBAbeld}
J_{\cal B}={1\over n}\mathscr{J}(\Phi)\,.
\fe
After finding this current, we can integrate out ${\frak b}$ and $c$ and use the simpler Lagrangian \eqref{btildeLag}.

The current \eqref{JBAbeld} shows that, the operator in \eqref{monopoleo}, which can be written as
\ie\label{monoAb}
{\cal O}=
 \Phi^{n}\prod_i{\cal M}_{{\frak b}^i}^{u_i}\,,
\fe
has charge $1$. 

After \eqref{monopoleo}, we emphasized that the factor of ${\cal M}_{\frak b}^n$ in $\cal O$ in \eqref{monopoleo}, means that $\cal O$ does not create fluxes for the modes of $\cal T$.  This is true in the presentation using the twisted fields ${\check b}^i$ in \eqref{frakbLagA}. However,the change of variables in \eqref{AbeKprime} means that the gauge fields ${\frak b}^i$ do have fluxes around $\cal O$ and these are represented by
$\prod_i{\cal M}_{{\frak b}^i}^{u_i}$ in \eqref{monoAb}.  As a check, the operator \eqref{monoAb} is invariant under all the gauge symmetries in the problem, including the $\U^r$ gauge fields ${\frak b}^i$.

Let us elaborate on the current \eqref{JBAbeld}.  As we stressed, it can be shifted by terms that vanish on-shell.  We will explore it using the variables in \eqref{btildeLag}.  The equations of motion of the $\U^r$ gauge fields ${\frak b}^i$ set $E^i= {1\over 2\pi } d{\frak b}^i -{u^i\over n} \mathscr{J}(\Phi)$ to zero. Therefore, on-shell, we can take the current to be
\ie
J_{\cal B}= {1\over n}\mathscr{J}(\Phi) +\sum_i \tau_iE^i={1\over 2\pi }\sum_i \tau_i  d{\frak b}^i +{1\over n}(1-\tau^\mathrm{T}u)\mathscr{J}(\Phi)
\fe
with arbitrary coefficients $\tau_i$.  Next, we choose special values of $\tau_i$.  Using \eqref{gcdnu}, we can find integer $\tau_i,{\cal Q_B}$ such that $\tau^\mathrm{T}u+n{\cal Q_B}=1$.  With these values, the on-shell current can be written as
\ie\label{currentint}
J_{\cal B}={1\over 2\pi }\sum_i \tau_i  d{\frak b}^i +{\cal Q_B} \mathscr{J}(\Phi)\,.
\fe
In this form, all the terms in the current lead to integer charges even off-shell.

This presentation of the current demonstrates a point we stressed in Section \ref{zeroforms}.  Even though in the form $J_{\cal B}={1\over n}\mathscr{J}(\Phi)$, the current includes fractional coefficients, all gauge-invariant operators have integer charges.

Another consequence of the presentation \eqref{currentint} is that turning on a background $\U$ gauge field $\cal B$ corresponds to adding to the Lagrangian $-{1\over 2\pi} \sum_i \tau_i {\cal B} d{\frak b}^i$ and replacing the kinetic term of $\Phi$ with $|D_{\sum_i q_i{\frak b}^i+{\cal Q_B}{\cal B}}\Phi|^2$, both with standard quantized coefficients.

\subsubsection{A dual description}

We can also use the dual description \eqref{dualLag} to express this theory as
\ie\label{dualAbea}
{1\over 4\pi} \sum_{i,j=1}^r{\frak b}^i  K_{ij}  d{\frak b}^j -{1\over 2\pi} \sum_{i=1}^r q_i {\frak b}^i d{\frak b}+ |D_{\frak b}\tilde \Phi|^2 +|d{\frak b}|^2-\tilde \mu^2|\tilde \Phi|^2 +\cdots\,.
\fe

For nonzero $p$, we can neglect the kinetic term of $\frak b$ and this is similar to \eqref{btildeLag}, except that 
\ie
&r\to r+1\\
&K\to \tilde K=K'=\begin{pmatrix}
        0&-q^\mathrm{T}\\-q&K
    \end{pmatrix}\\
&q\to \tilde q=\begin{pmatrix}1\\0\end{pmatrix}\\
&\mu^2\to\tilde \mu^2=-\mu^2\,,
\fe
where in $\tilde q$, $0$ is an $r$-dimensional vector. Note that $K'$ is as in \eqref{Krp}. Using this data, we find
\ie\label{tildeKq}
        &\det \tilde K=-q^{\mathrm T} K^{-1}q \det K =-{p\over n}\det K\\
       &\tilde u=\tilde n \tilde K^{-1}\tilde q =\begin{pmatrix}
            -n\\u
            \end{pmatrix}\\ &\tilde n=p\\
            &\tilde p=\tilde n\tilde q^{\mathrm T}\tilde K^{-1}\tilde q=-n\,.
    \fe
(For simplicity, we assumed that $p>0$.) As a check, this map of $n$ and $p$ to $\tilde n$ and $\tilde p$ is as in Section \ref{inversemap}.

This leads to the $\U_{\cal B}$ current
\ie\label{}
&J_{\cal B}= {1\over 2\pi n}  d{\frak b}=-{1 \over p} \mathscr{J}(\tilde \Phi)\\
&\mathscr{J}(\tilde \Phi)=\star i\tilde\Phi\overset\leftrightarrow{D} \bar{\tilde\Phi}\,,
\fe
and to the operator \eqref{monoAb} as
\ie\label{OAbed}
{\cal O}=\left(\bar{\tilde\Phi}\right)^{p}{\cal M}_{\frak b}^n\prod_i{\cal M}_{{\frak b}^i}^{u_i}\,.
\fe

One might be puzzled by the fact that the expression for $\cal O$ in \eqref{monopoledual} includes only $\left(\bar{\tilde\Phi}\right)^{p}{\cal M}_{\frak b}^n$ without the second factor of $\prod_i{\cal M}_{{\frak b}^i}^{u_i}$. The reason for that is related to our comments after \eqref{monopoleo} and \eqref{monoAb}. Here, we specified the induced fluxes of ${\frak b}^i$ and ${\frak b}$, while the meaning of \eqref{monopoledual}, in our context, is that the fluxes of $\check  b^i$ around $\cal O$ vanish.

These expressions clearly demonstrate the more general analysis in Section \ref{dual theory}.

\subsubsection{Coupling the theory to a background $\U_A$ gauge field}\label{Abenrichement}

As above, we start by enriching $\cal T$.  We do that by using a vison $v$.  In the variables in \eqref{TAbeLag} we write the vison as $\exp\left(i\sum_i t_i\int b^i\right)$.  This corresponds to coupling the Lagrangian \eqref{TAbeLag} to a background $\U_A$ gauge field $A$ by adding to it
\ie\label{Atbcoupling}
{1\over 2\pi} A\sum_i t_i db^i\,.
\fe

Using the notation in Section \ref{Enrichment}, 
\ie
n_v=t^\mathrm{T}u\bmod n\,,
\fe
and the $\U_A$ charge of $a$ is
\ie
Q_A(a)={n_v\over n}\bmod 1\,.
\fe
This leads to the response theory $-{t^\mathrm{T}K^{-1} t\over 4\pi}  AdA$.  Taking a possible counterterm into account, the TQFT $\cal T$ has Hall conductivity
\ie
\sigma_H^A=t^\mathrm{T}K^{-1} t\bmod 2\,.
\fe

Next, we study the transition theory. The Lagrangian \eqref{btildeLagp} or alternatively \eqref{btildeLag}, is coupled to $A$ as 
\ie\label{btildeLagA}
{1\over 4\pi} \sum_{i,j=1}^r {\frak b}^i K_{ij}d{\frak b}^j+{1\over 2\pi} \sum_i t_i {\frak b}^i dA+ |D_{\sum_i q_i{\frak b}^i+{\cal Q}A}\Phi|^2 -\mu^2|\Phi|^2 + \cdots\,,
\fe
where we allowed a direct coupling of $\Phi$ to $A$ based on the integer charge $\cal Q$.

This coupling to $A$ guarantees that for positive $\mu^2$ we end up with $\cal T$ enriched as above with 
\ie
\sigma_H^A(\mu^2>0)=t^\mathrm{T}K^{-1} t\bmod 2\,.
\fe

For negative $\mu^2$, $\Phi$ Higgses the gauge group setting $\sum_i q_i{\frak b}^i+{\cal Q}A $ to a pure gauge.  This can be implemented by adding a Lagrange multiplier $b^0$ and studying the Lagrangian 
\ie
&{1\over 4\pi} \sum_{i,j=1}^r {\frak b}^i K_{ij}d{\frak b}^j-{1\over 2\pi}b^0\sum_{i=1}^r q_id{\frak b}^i+ {1\over 2\pi}\left( \sum_{i=1}^r t_i {\frak b}^i -{\cal Q} b^0\right)dA \,.
\fe
 Thus, the K-matrix becomes $ K'$ of \eqref{Krp} and the charge vector becomes
\ie
  t'= \begin{pmatrix}-{\cal Q}\\ t\end{pmatrix}\,.   
\fe
It leads to the Hall conductivity
\ie
\sigma_H^A(\mu^2<0)&={ (t')}^\mathrm{T} { (K')}^{-1} { t'} \bmod 2\\
&= \left(t^\mathrm{T}K^{-1}t - \frac{(t^\mathrm{T}u-n\mathcal{Q})^2}{np}\right)
\bmod 2.
\fe

Remembering that both sides of the transition have the same ${1\over 2\pi}AdA$ counterterm, these expressions are consistent with \eqref{changesigma}
\ie\label{changesigmas}
&\sigma_H^A(\mu^2<0)=\sigma_H^A(\mu^2>0)-{n_v^2\over pn} \\
&n_v= t^\mathrm{T}u -n{\cal Q}\\
&J_A={n_v\over n} \mathscr{J}(\Phi)\,.
\fe

The discussion of this example with $p=0$, is identical to that in the general theory in Section \ref{piszero}.  The explicit expressions in the concrete example of the Abelian theory are identical to those in the general theory.  Therefore, we will not repeat them here.

\subsubsection{A special case}

As a more concrete example, consider the ${\cal T}=\U_8\to{\cal T}'= \bZ_4$ transition.

The $\U_8$ TQFT has a $\bZ_8^{(1)}$ symmetry, which is generated by the anyon $a_1$.  Its anomaly is $1$.  We focus on the anyon $a=a_1^4$, which generates $\bZ_2^{(1)}$ (hence, $n=2$), with anomaly $p=0\bmod 4$.  

If we take $p=0$, we find the gauging transition in Section \ref{piszero}, and we end up with ${\cal T}'=\U_2$.

Instead, we consider $a=a_1^4$ with $p=4$.  This leads to ${\cal T}'={\U_8\boxtimes\U_{-8}\over \bZ_2}$, which is the $\bZ_4$ gauge theory.  Following \eqref{btildeLag} and denoting the gauge field ${\frak b}^1$ as $b_e$, the transition theory is
\ie\label{U1toZ4e}
{8\over 4\pi}b_edb_e+|D_{4 b_e}\Phi|^2 -\mu^2|\Phi|^2 + \cdots\,.
\fe
The coupling to $\Phi$ breaks the $\bZ_8^{(1)}$ symmetry with anomaly $1$, to a $\bZ_4^{(1)}$ symmetry with anomaly $2\bmod 8$, which is generated by $a_1^2$.

In the form \eqref{U1toZ4e}, it is clear that the transition Higgses the $\U$ gauge theory to a $\bZ_4$ gauge theory.

Using  \eqref{dualAbea}, the dual theory is
\ie\label{}
&{8\over 4\pi} b_edb_e-{4\over 2\pi}  b_ed{\frak b}+ |D_{\frak b}\tilde \Phi|^2 +\cdots={4\over 2\pi} b_edb_m + |D_{b_e-b_m}\tilde \Phi|^2 +\cdots\\
&b_m=b_e-{\frak b}\,.
\fe
The first term represents the $\bZ_4$ gauge theory \cite{Maldacena:2001ss,Banks:2010zn,Kapustin:2014gua}.

The pure $\bZ_4$ gauge theory has a $\bZ_4^{(1)}\times \bZ_4^{(1)}$ symmetry generated by $e$ and $m$.  The coupling of $\tilde \Phi$ to the gauge fields shows that $\tilde \Phi$ corresponds to the anyon $e\bar m$. This coupling breaks the $\bZ_4^{(1)}\times \bZ_4^{(1)}$ symmetry of the pure gauge theory to the diagonal $\bZ_4^{(1)}$, which is generated by $em$.  Its anomaly is $2\bmod 8$.   As a check, the original theory and its dual have the same one-form symmetry and the same anomaly.  Note that in this case, the anyon created by $\tilde \Phi$, $e\bar m$, is not included in the one-form symmetry.

The symmetry anyons are present on both sides of the transition.  They are mapped as
\ie
\U_8 &\qquad &\bZ_4\\
\hline
1\quad&\quad\longleftrightarrow &1\\
a_1^2 \quad&\quad\longleftrightarrow &em\\
a_1^4\quad& \quad\longleftrightarrow &e^2m^2\\
\bar a_1^2\quad &\quad\longleftrightarrow& \bar e\bar m
\fe
Starting in the $\U_8$ side of the transition, the transition is triggered by the proliferation of $a=a_1^4$, which is one of the symmetry lines.  In the $\bZ_4$ side of the transition, this line is interpreted as $e^2m^2$.  From this side, the transition is triggered by the proliferation of $e\bar m$.  It does not correspond to any well-defined anyon on the $\U_8$ side of the transition.  Interestingly, $\tilde \Phi^2$ creates the anyon $e^2m^2$, which has the same quantum numbers as the anyon created by $\Phi$ on the other side of the transition.

\subsubsection{Jain states}

Let us apply our theory to possible transitions between  Jain states \cite{JainCF,jain2007composite}.

The Jain series TQFT ${\cal J}_{r,s}$  can be described by an Abelian theory with $\U^r$ gauge fields $b^i$ \cite{wen1995topological}  
\ie\label{JaiLag}
 &{\cal L}_{{\cal J}_{r,s}}= \frac{1}{4\pi}\sum_{i,j=1}^rb^i K_{ij}(r,s)db^j + \frac{1}{2\pi}A\sum_{i=1}^r t_i db^i\\
&K_{ij}(r,s)=\begin{cases}
        s & i\neq j\\
        s+1 & i=j
        \end{cases}
        \qquad, \qquad  t_i=1\,.
        \fe
For odd $s$, this is a bosonic theory and $A$ is a background $\U$ gauge field.  For even $s$, this is a spin$^c$ theory and $A$ is a spin$^c$ connection.  Since we focus on bosonic theories, we will consider odd $s$, and, for simplicity, we will take $s>0$.

Importantly, ${\cal J}_{r,s}$ is a minimal TQFT ${\cal A}^{n,p}$ \cite{JensenRaz,Cheng:2025ube}. (See Section \ref{calT}.)  The value of $n$ is 
\ie
n=\det K(r,s)=1+rs\,,
\fe
but the value of $p\bmod 2n$ depends on the anyon we take as the generator of the $\bZ_n^{(1)}$ symmetry \cite{Hsin:2018vcg}.

For example, if we take the anyon $ a=e^{i\int b^r} $ (denoted $q_i=\delta_{i,r}$) as the generator, 
$q^\mathrm{T}K(r,s)^{-1} q={\det K(r-1,s)\over \det K(r,s)}={1+(r-1)s\over 1+rs}$ 
and 
\ie\label{generq}
p=(1+(r-1)s)\bmod 2n \qquad, \qquad a=e^{i\int b^r}\,.
\fe
Alternatively, if we take the vison $v=e^{i\sum_i\int b^i}$ as the generator (see the coupling to $A$ in \eqref{JaiLag}), 
\ie\label{generv}
p=r\bmod 2n\qquad, \qquad v=e^{i\sum_i\int b^i}\,.
\fe
These different generators lead to different presentations of ${\cal J}_{r,s}$
\ie\label{JisA}
{\cal J}_{r,s}\cong {\cal A}^{1+rs,1+(r-1)s} \cong {\cal A}^{1+rs,r}\,.
\fe

Next, we would like to study transitions away from the TQFT ${\cal J}_{r,s}$.  This depends on the choice of anyon $a$ that triggers the transition. If $a$ becomes light at the transition, so do all its powers. In our case, these are all the anyons in ${\cal J}_{r,s}$. The resulting transition theory and the theory on the other side of the transition $\cal T'$ depend on which anyon we focus on.  Presumably, this is the lightest nontrivial anyon.  Here, we will focus on  $a=e^{i\int b^r}$.  (If we focus on another anyon, e.g., the vison \eqref{generv}, we would end up with a different theory.)

In addition, the transition theory depends on the choice of $p$ with a given $p\bmod 2n$.  First, we will study $p=-1-(r+1)s$.  As in footnote \ref{Lisonef}, this leads to the map
\ie
{\cal J}_{r,s} \to {{\cal J}_{r,s}\boxtimes \U_{(1+rs)(1+(r+1)s)}\over \bZ_{1+rs}} \cong {\cal J}_{r+1,s} \,.
\fe
The inverse map, which is related to the dual of the transition theory, is 
\ie
 {\cal J}_{r+1,s}\to {{\cal J}_{r+1,s}\boxtimes \U_{-(1+(r+1)s)(1+rs)}\over \bZ_{1+(r+1)s}}\cong {\cal J}_{r,s}\,.
\fe
This inverse map, which reduces $r$ is related to a transition from ${\cal J}_{r,s}$ with the same choice of $a$, but with $p=1+(r-1)s$
\ie\label{changeingr}
 {\cal J}_{r,s}\to {{\cal J}_{r,s}\boxtimes \U_{-(1+rs)(1+(r-1)s)}\over \bZ_{1+rs}}\cong {\cal J}_{r-1,s}\,.
\fe

These maps show that our transition theories can map $r\to r\pm 1$.  Additional transitions, including transitions that take us outside the Jain series ${\cal J}_{r,s}$, can be found by keeping the same anyon $a$ and shifting $p$ by a multiple of $2n$.  And as we said above, we can also choose other anyons in ${\cal J}_{r,s}$ (e.g., the vison $v$) with different $p\bmod 2n$, and find more transitions to other TQFTs.

Next, we discuss the Hall conductivity.  The coupling to $A$ in \eqref{JaiLag} identifies the vison as the anyon with $t_i=1$.  From this, we find
\ie
n_v=n t^{\mathrm T}K^{-1}q \bmod n =1\bmod n\,.
\fe
The Hall conductivity is
\ie
\sigma_H({\cal J}_{r,s})=t^{\mathrm T}K^{-1}t={r\over 1+rs}={r\over n}\,,
\fe
and therefore, the change across the transition corresponding to \eqref{changeingr} is
\ie
\sigma_H({\cal J}_{r,s})-\sigma_H({\cal J}_{r-1,s})={1
\over (1+rs)(1+(r-1)s)}={n_v^2\over np}\,,
\fe
where we took $n_v=1$.  This is consistent with our general result \eqref{changesigma} and the specialization to the Abelian theory \eqref{changesigmas}.

\subsection{${\cal T}={\rm SU}(2)_k$}\label{TisSU2}

\subsubsection{$\cal T$ and ${\cal T}'$}\label{TTpSU2}

Here, we demonstrate the general discussion in the case of the non-Abelian theory
${\cal T}=\rm{SU}(2)_k$. It has a $\bZ_2^{(1)}$ one-form symmetry generated by the order $n=2$ anyon $a$ with $j={k\over 2}$ and $p=k\bmod 4$.

As above, we study
\ie\label{TTps}
&{\cal T}'={{\rm SU}(2)_k\boxtimes \U_{-2p}\over \bZ_2}={{\rm SU}(2)_k\boxtimes \U_{-2p}\over (j={k\over 2},p)}\\
&p=k-4N\qquad, \qquad k,N\in \bZ\,
\fe
and describe $\cal T$ and $\cal T'$ using a ${\rm U}(2)$ gauge field $\frak a$ and a $\U$ gauge field $c$ with the Lagrangians \cite{seiberg2016gapped,Hsin:2016blu}
\ie\label{TprimeSU2}
&\mathcal{L}_{{\cal T}}=\frac{k}{4\pi}\Tr \Big({\frak a}d{\frak a}-\frac{2i}{3} {\frak a}^3\Big) -\frac{k}{4\pi}(\Tr {\frak a})d(\Tr{\frak a})+{1\over 2\pi} cd(\Tr\frak{a})\\
&\mathcal{L}_{{\cal T}'}=\frac{k}{4\pi}\Tr \Big({\frak a}d{\frak a}-\frac{2i}{3} {\frak a}^3\Big) -\frac{k-2N}{4\pi}(\Tr {\frak a})d(\Tr{\frak a})\,.
\fe
In the Lagrangian for $\cal T$,  $c$ sets $\Tr \frak{a}$ to a trivial gauge field, thus making the theory an SU(2) theory.

To relate this discussion to the general analysis above, we decompose the $\rm{U}(2)$ gauge field as\footnote{The U(2) field $\frak a$ here and the U(1) fields ${\frak b}^i$ and $\frak b$ in the Abelian discussion in Section \ref{Abelianexamples} are denoted by the same font as $\frak b$ in our general discussion.  They play similar roles.  See also footnote \ref{differentbs}.}
\ie\label{SU2deco}
&{\frak a}={\check  b}\sigma^0+\sum_i {\check  b}^i\sigma^i\\
&\Tr {\frak a}=2{\check  b}={\frak b}\,.
\fe
Here, $\sigma^i$ are the Pauli matrices and $\sigma^0$ is the identity matrix. Without the quotient in \eqref{TTps}, ${\check  b^i}$ and $\check  b$ would have been standard ${\rm SU}(2)$ and $\U$ gauge fields, respectively.  The quotient makes them twisted fields.  ${\check  b}^i$ are $\rm{SO}(3)$ gauge fields and, as above, $\check  b$ is a twisted $\U$ gauge field. ${\frak b}=\Tr {\frak a}$ is a standard $\U$ gauge field, whose fluxes are correlated with the second Stiefel–Whitney class
$w_2$ of the ${\rm SO}(3)$ gauge fields ${\check b}^i$ as  $w_2= {1\over 2\pi}\int_{{\cal C}_2} d(\Tr\frak {a})\bmod 2$.

As in Section \ref{Abelianexamples}, the fact that we have an explicit Lagrangian for $\cal T$ allowed us to have a presentation of the theory using standard gauge fields $\frak a$ and $c$, rather than the twisted fields $\check  {b}^i$ and $\check  b$.

The line operator $\exp\left(i\int_{{\cal C}_1} c\right)$ generates the $\bZ_2^{(1)}$ symmetry of $\cal T$.  One way to see that is to note that an insertion of that line leads to $d\,(\Tr\frak {a})=-2\pi \delta({\cal C}_1)$.  If ${\cal C}_1$ is topologically nontrivial, it shifts the dual $w_2$ co-cycle.  It also has the effect of multiplying any line operator with SU(2) quantum number $j$ that winds around ${\cal C}_1$ by $(-1)^{2j}$.  This is the action of $\bZ_2^{(1)}$.

The Lagrangians \eqref{TprimeSU2} demonstrate one of the points in Section \ref{TTprel}. In this description of $\cal L_T$, it is easy to couple it to a background U(1) gauge field $X$ and add an SPT (a counterterm) ${N\over 2\pi}XdX$
\ie\label{TprimeSU2E}
\mathcal{L}_{{\cal T}}=\frac{k}{4\pi}\Tr \Big({\frak a}d{\frak a}-\frac{2i}{3} {\frak a}^3\Big) -\frac{k}{4\pi}(\Tr {\frak a})d(\Tr{\frak a})+{1\over 2\pi} c(d(\Tr\frak{a})+dX)+{N\over 2\pi}XdX\,.
\fe
As we said above, the line operator $\exp\left(i\int c\right)$ generates the $\bZ_2^{(1)}$ symmetry of $\cal T$.  Therefore, we recognize $\left[{dX\over 2\pi}\right]_2$ as its background gauge field.  And the coupling to $X$ enriches $\cal T$ by $\U_X$ via its $\bZ_2^{(1)}$ symmetry.\footnote{The equations of motion of ${\frak a}$ and $c$ state that 
\ie\label{aceomwithX}
&dc=-{k\over 2}dX\\
&d(\Tr {\frak a})=-dX\,.
\fe
For odd $\int_{{\cal C}_2}{dX\over 2\pi}$,  $\int_{{\cal C}_2}{d(\Tr {\frak a})\over 2\pi}$ should also be odd, corresponding to nonzero $w_2$.  In this case, if $k$ is also odd, these equations can be solved only if we add delta-function sources with nontrivial holonomy of $c$ around them.  The minimal such addition is a single line associated with the special anyon $a$, i.e., $e^{i\int c}$.  This line shifts the flux of $c$ by one unit and enables a solution of \eqref{aceomwithX}.\label{aceomSU2}}

If we turn $X$ into a dynamical field, $x$, it is easy to integrate out $x$ and $c$, and end up with $\cal L_{T'}$ in \eqref{TprimeSU2}.  This is the relation between $\cal T$ and $\cal T'$ in Section \ref{Cppgau}.

\subsubsection{The non-topological transition field theory}\label{SU2UVtheory}

Next, we add $\Phi$ and study the Lagrangian \eqref{Higgstheory}
\ie\label{twPhisu2}
\mathcal{L}=\mathcal{L}_{{\cal T}'}+ \frac{1}{2\pi}c d(\Tr {\frak a}) + |D_c\Phi|^2-\mu^2|\Phi|^2+\cdots\,.
\fe

For $p=k-4N\ne 0$, this leads to a transition
\ie\label{SU2quo}
    {\rm SU}(2)_k \to {{\rm SU}(2)_k\boxtimes \U_{-2(k-4N)}\over \bZ_2} ={{\rm SU}(2)_k\boxtimes \U_{-2(k-4N)}\over(j={k\over 2},k-4N)}\,.
\fe
The special case of $N=0$ describes the transition from  ${\rm SU}(2)_k$ to ${{\rm SU}(2)_k\boxtimes \U_{-2k}\over \bZ_2}$.  The latter theory is the TQFT counterpart \cite{Moore:1989yh,seiberg2016gapped} of the GKO construction  ${{\rm SU}(2)_k\over \U_{2k}} $ of the $\bZ_k$ parafermion theory \cite{Goddard:1984vk,Goddard:1986ee}.

For $k=4N$ (and hence $p=0$), the transition is to a gapless theory.  Adding the operator\footnote{As we discussed in various places (following \eqref{monopoleo},  \eqref{monoAb}, and  \eqref{OAbed}), the meaning of ${\cal M}_{\Tr \frak{a}}^2$ is that the field $\Tr \frak{a}$ has $4\pi$ flux around $\cal O$, and the fields $\check {b}^i$ are smooth there.} \eqref{monopoleo} ${\cal O}= \Phi^{2}{\cal M}_{\Tr \frak{a}}^2$ to the Lagrangian leads to a transition from ${\rm SU}(2)_{4N}$ to ${\rm SO}(3)_{4N}$.\footnote{There are several different definitions of ${\rm SO}(3)_k$.  Here, we use ${\rm SO}(3)_k={{\rm SU(2)}_k\over \bZ_2}$.}

\subsubsection{$\U_{\cal B}$}\label{SU2B}

It is interesting to compare this example with the Abelian one in Section \ref{UoneB}.  There, we showed that on-shell, ${1\over 2\pi}\int_{{\cal C}_2}d{\frak b}\in n\bZ$, i.e., $\ell=1$.   This is not the case here.

The equations of motion of the SO(3) fields ${\check b}^i$ set the SO(3) field strength to zero, but they do not restrict its $w_2$.  Therefore, they do not restrict the fluxes of ${\frak b}=\Tr\frak{a}$.  Consequently, the only restriction is due to the equation of motion of ${\frak b}$ \eqref{restfrombt}
\ie
{1\over 2\pi }\int_{{\cal C}_2}d{\frak b}={1\over 2\pi }\int_{{\cal C}_2}d(\Tr {\frak a})\in {n\over L}\bZ={2\over \gcd(2,k)}\bZ\,,
\fe
corresponding to $\ell=L=\gcd(2,k)$.  Therefore, the properly normalized conserved current is
\ie
J_{\cal B}={\gcd(2,k)\over 2}\mathscr{J}(\Phi)={\gcd(2,k)\over 4\pi}d(\Tr\frak{a})={m_b\over 2\pi}  d{(\Tr\frak{a})}+{m_c\over 2\pi}  d{c}\,,
\fe
 with $m_b$ and $m_c$ integer solutions of $2m_b+pm_c=\gcd(2,k)$.
(See \eqref{secondatt}.)  The last form of the current makes it manifest that its charges are quantized.

We learn that for odd $k$, the operator ${\cal O}=\Phi^2{\cal M}_{\Tr\frak{a}}^2{\cal M}_c^{p}$ has charge 1, and for even $k$, it has charge 2.

Let us examine it in more detail.  For odd $k$, the on-shell fluxes are constrained to satisfy ${1\over 2\pi}\int_{{\cal C}_2} d(\Tr {\frak a})\in 2\bZ$ and this means that the on-shell SO(3) bundles have vanishing $w_2$.

For even $k$, we can also have ${1\over 2\pi}\int_{{\cal C}_2} d(\Tr {\frak a})\in 2\bZ+1$, which corresponds to nontrivial $w_2$.  In general, this is consistent with the fact that the SO(3) field strength vanishes.  However, this is not possible when ${\cal C}_2$ is a sphere.  The local operators of the theory correspond to states on the sphere.  Therefore, this argument shows that for even $k$, the local operators must have ${1\over 2\pi}\int_{{\cal C}_2} d(\Tr{\frak a})\in 2\bZ$ and hence, even $\U_{\cal B}$ charges.  Consequently, the operator ${\cal O}'=\Phi{\cal M}_{\Tr\frak{a}}{\cal M}_c^{p\over 2}$, which is associated with odd $\Tr {\frak a}$ flux, and therefore nonzero $w_2$, is not a local operator. It can, however, act on the torus. And when it does, it changes the eigenvalue of $\bZ_2^{(0)}\subset \U_{\cal B}$.

 Let us describe more concretely the states on the torus that are $\bZ_2^{(0)}$ odd. Following the discussions in Section \ref{zeroforms}, when $\Tr \frak{a}$ is treated as a classical background field, the torus states with $\frac{1}{2\pi}\int d (\Tr\frak{a})\in 2\bZ+1$ have SO(3) bundles with nonzero $w_2$. In the TQFT ${\cal T}={\rm SU(2)}_k$, this background $\Tr\frak{a}$ corresponds to a line with isospin $j={k\over 4}$ running through the interior of the torus and a defect with $j={k\over 2}$ piercing the torus \cite{Moore:1989yh}.  This configuration is special because of the fusion rule ${k\over 4} \times {k\over 2}={k\over 4}$. Restoring $\Tr\frak{a}$ as a dynamical field, and embedding it in the full transition theory, this can be thought of as acting with ${\cal O}'$ on the torus with a $\bZ_2^{(0)}$ even state.

The conclusion is that for even $k$, the $\U_{\cal B}$ symmetry includes a $\bZ_2^{(0)}$ subgroup that acts only on line-operators, but not on local operators.

\subsubsection{The states on the torus for $p=0$}\label{SU2ellistwo}

Here, we focus on the special case of $p=0$, and therefore $k=0\bmod 4$. Before deforming the theory by $\cal O$, the negative $\mu^2$ phase is gapless.  Let us examine the low-energy spectrum on the torus in that phase. In particular, we would like to see the consequences of the fact that $\ell=2$.

Following Section \ref{simplifyingt}, we write the theory using ${\mathbf c}^{(1)}$ and ${\mathbf g}^{(2)}$.  Then we can simplify the discussion by first setting one of them to zero to find two decoupled theories, and then later restoring the gauge field.  We can do this for any of these gauge fields.

Let us set  ${\mathbf c}^{(1)}=0$.  Then, for negative $\mu^2$, we have two decoupled sectors SO(3)$_k$ and a compact boson (denoted by $\alpha$ in Section \ref{piszerop}).  Ignoring the oscillator modes of the compact bosons, the states on the torus are
\ie
&{1\over \sqrt{2}}\left(|j, Q, w_x,w_y\rangle +|{k\over 2}-j, Q, w_x,w_y\rangle\right)\quad, \quad j=0,1,\cdots,{k\over 4}-1\\
&|j={k\over 4}, Q, w_x,w_y\rangle \\
&|j=\left({k\over 4}\right)', Q, w_x,w_y\rangle\\
&Q,w_x,w_y\in \bZ\,.
\fe
Here, we label the states by the corresponding states in the SU(2)$_k$ theory.  The state labeled by $j$ corresponds to a Wilson line with $j$ in the interior of the torus along the $x$ direction, and $|j=\left({k\over 4}\right)', Q, w_x,w_y\rangle$ is from the twisted sector of SO(3)$_k$.  Comparing with Section \ref{piszerop}, the winding numbers $w_i\in \bZ$ with $i=x,y$ differ from the winding there by a factor of $2$.  (See footnote \ref{windingnor}).  This factor of $2$ will be clarified below.  And the $\U$ ``momentum charge'' $Q$ is the charge of $\U_{\cal B}$.

On a torus with sides $L_x\times L_y$, the energy of these states is of order
\ie\label{energyQwi}
{Q^2\over L_xL_y}+ {L_yw_x^2\over L_x}+ {L_xw_y^2\over L_y}\,,
\fe
where we suppressed coefficients.  Importantly, for a large torus, the states with $w_i=0$ are light, but the winding modes have energy of order one.

In the full theory (still without ${\mathbf c}^{(1)}$), the $\U^{(1)}$ winding symmetry is emergent and therefore the winding numbers $w_i$ are not meaningful.  And since the states with nonzero winding have energy of order one, we can ignore $w_i$.  We keep these labels because soon they will be important.

Next, we restore the coupling to ${\mathbf c}^{(1)}$.  It gauges the $\bZ_2^{(0)}$ symmetry that acts on the compact boson as $(-1)^Q$ combined with the symmetry that assigns a minus sign to  $|j=\left({k\over 4}\right)', Q, w_x,w_y\rangle$. See the discussion around \eqref{symmetryg}.

The states in the untwisted sector are
\ie\label{untwisted}
&{1\over \sqrt{2}}\left(|j, Q, w_x,w_y\rangle +|{k\over 2}-j, Q, w_x,w_y\rangle\right)\quad , \quad Q\in2\bZ\quad, \quad j=0,1,\cdots,{k\over 4}-1\\
&|j={k\over 4}, Q, w_x,w_y\rangle \qquad , \qquad Q\in2\bZ\\
&|j=\left({k\over 4}\right)', Q, w_x,w_y\rangle\qquad , \qquad Q\in2\bZ+1\\
&w_x,w_y\in \bZ\,.
\fe

For a large torus, the light states have $w_x=w_y=0$. In this sector, the states with even and odd $Q$ have different degeneracies for fixed $Q$.  Ignoring the degeneracy between the states with $\pm Q$, there are $k/4+1$ states for even $Q$, and one state for odd $Q$. 

There are three kinds of twisted sector states.  Those that are twisted around the $x$ direction are
\ie\label{twistedx}
&{1\over \sqrt{2}}\left(|j, Q, w_x,w_y\rangle -|{k\over 2}-j, Q, w_x,w_y\rangle\right)\\
& Q\in2\bZ\quad, \quad j=0,1,\cdots,{k\over 4}-1\quad,\quad w_x\in \bZ+{1\over 2}\quad, \quad w_y\in \bZ\,.
\fe
Those that are twisted around the $y$ direction are
\ie\label{twistedy}
&{1\over \sqrt{2}}\left(|j, Q, w_x,w_y\rangle +|{k\over 2}-j, Q, w_x,w_y\rangle\right)\\
& Q\in2\bZ\quad, \quad j={1\over 2},{3\over 2},\cdots,{k-2\over 4}\quad, \quad w_x\in \bZ\quad, \quad w_y\in \bZ+{1\over 2}\,.
\fe
And those that are twisted both around $x$ and around $y$ are
\ie\label{twistedxy}
&{1\over \sqrt{2}}\left(|j, Q, w_x,w_y\rangle -|{k\over 2}-j, Q, w_x,w_y\rangle\right)\\
& Q\in2\bZ\quad, \quad j={1\over 2},{3\over 2},\cdots,{k-2\over 4}\quad, \quad w_x,w_y\in \bZ+{1\over 2}\,.
\fe

These expressions for the spectrum demonstrate our general discussion in Section \ref{piszerop} of the mixing between the Goldstone mode $\alpha$ and the modes in the TQFT ${\cal T}={\rm SU(2)}_k$.  While comparing with Section \ref{piszerop}, note that the winding numbers there are $2w_i$ (see footnote \ref{windingnor}) and the charge $Q$ is normalized as the charge of $\U_{\cal B}$ there.

Gauging ${\mathbf c}^{(1)}$ leads to a dual $\bZ_2^{(1)}$ symmetry that acts as $(-1)^{2w_i}=W_i(j={k\over 2})$  with $W_i(j={k\over 2})$ the SU(2) Wilson line with $j={k\over 2}$ along direction $i$.  This is the $\bZ_2^{(1)}$ symmetry of our transition theory.

Alternatively, consider first ignoring ${\mathbf g}^{(2)}$, but keeping ${\mathbf c}^{(1)}$.  This means that the states are
\ie
|j,Q,w_x,w_y\rangle \quad, \quad j=0,{1\over 2},\cdots ,{k\over 2} \quad,\quad Q\in 2\bZ\quad, \quad w_i\in {1\over 2}\bZ\,,
\fe
where we followed the previous normalizations. Next, we restore ${\mathbf g}^{(2)}$ by gauging the $\bZ_2^{(1)}$ generated by  $W_i(j={k\over 2})(-1)^{2w_i}$. In the untwisted sector, we find all the states in \eqref{untwisted} except those with $\left({k\over 4}\right)'$, as well as the states in \eqref{twistedx}, \eqref{twistedy}, \eqref{twistedxy}.  As a check, they are invariant under $W_i(j={k\over 2})(-1)^{2w_i}$.  In the twisted sector, we find the states in \eqref{untwisted} with $\left({k\over 4}\right)'$ and odd $Q$.

As we said above, in the full theory, $w_i$ are not meaningful.  However, $w_i\bmod 1$ are meaningful and $(-1)^{2w_i}$ are the eigenvalues of the $\bZ_2^{(1)}$ symmetry of the transition theory.  This symmetry is unbroken.  Indeed, the charged states have energy of order one.  See \eqref{energyQwi}.

Finally, as in our general discussion, the $\U_{\cal B}$ charges $Q$ of all the local operators are even, in accordance with $\ell=2$.  There are, however, states on the torus with odd $Q$.  They are related to the other states using the line operator ${\cal O}'$, which has $Q=1$.

\subsubsection{A dual description}

As in \eqref{dualLag}, the dual Lagrangian is
\ie\label{dualLags}
\frac{k}{4\pi}\Tr \Big({\frak a} d{\frak a} -\frac{2i}{3} {\frak a} ^3\Big) -\frac{k-2N}{4\pi}(\Tr{\frak a} )d(\Tr {\frak a} )   +|D_{ \Tr {\frak a} }\tilde\Phi|^2  +|d(\Tr {\frak a})|^2+\mu^2|\tilde \Phi|^2 + \cdots\,.
\fe
Here, it is clear that for $\mu^2<0$, $\tilde \Phi$ is massive and we find a gapped phase with ${\cal T}'={{\rm SU}(2)_k\boxtimes \U_{-2(k-4N)}\over \bZ_2}$.  (For $k=4N$ this theory is gapless.)  And for $\mu^2>0$, $\tilde \Phi$ Higgses $\Tr{\frak a}$, and we end up with a gapped theory with ${\cal T}={\rm SU}(2)_k$.

Approaching the transition from the positive $\mu^2$ side, it is triggered by the anyon $a$, corresponding to the SU(2) representation $j={k\over 2}$.  The dual picture \eqref{dualLags} shows that approaching the transition from the negative $\mu^2$ side, the anyon that triggers it is SU(2) invariant and with U(1) charge $2$.  (As a check, this anyon is consistent with the quotient in \eqref{SU2quo}.)

\subsubsection{Coupling the theory to a background $\U_A$ gauge field}\label{SU2enrichement}

Here, the starting point, ${\cal T}={\rm SU}(2)_k$, has only a $\bZ_2^{(1)}$ symmetry, generated by $a$.  Hence, there is only one nontrivial way to enrich it by $\U_A$, using the vison $v=a$, as we did in \eqref{TprimeSU2E}.  As a result,
\ie\label{QAaSU2}
&Q_A(a)=\langle a,a\rangle={k\over 2}\bmod 1\\
&n_v=k\bmod 2\,.
\fe

If $n_v=0$, which is possible only for even $k$, $J_A$ does not act faithfully in the transition theory, and the coupling of $A$ is a pure enrichment.  In this case, the existence of the  $\U_A$ symmetry does not affect the discussion of the behavior of the $p=0$ theory.  

If $k$ is even and $n_v\ne 0$,  $\U_A$ acts faithfully in the transition theory and $Q_A({\cal O})=n_v\ne 0$.  As a result, for $p=0$, we cannot deform the theory by $\cal O$ and for negative $\mu^2$, $\U_A$ is spontaneously broken.  However, as in footnote \ref{UVcalQ}, depending on how this theory arises from a UV theory, we may be able to change $\cal Q$, set $n_v=0$, deform the theory by $\cal O$, and gap the negative $\mu^2$ phase.

For odd $k$, the fractional $Q_A(a)$ in \eqref{QAaSU2} means that $J_A$ acts faithfully in the transition theory.  It is not a pure enrichment coupling.  In particular, $Q_A({\cal O})=n_v\ne 0$. However, since for odd $k$, we cannot have $p=0$,  the subtleties with fractional $Q_A(a)$ for $p=0$ do not arise in this case of ${\cal T}={\rm SU}(2)_k$.

More conceptually, for odd $k$, the coupling of $\Phi$ breaks the $\bZ_2^{(1)}$ symmetry generated by $v=a$, and therefore, the coupling of $\U_A$ to the theory cannot be a pure enrichment coupling.

Let us reach the same conclusion using the Lagrangian.  We would like to couple \eqref{twPhisu2} to $A$ such that at the level of the TQFT $\cal T$, it is the same as \eqref{TprimeSU2E} with $A=X$.  For simplicity, we neglect the freedom to add an $AdA$ counterterm.  Clearly, we should write
\ie\label{SU2ene}
\mathcal{L}=&\mathcal{L}_{{\cal T}'}+ \frac{1}{2\pi}c \big(d(\Tr {\frak a})+dA\big) + |D_{c+{\cal Q}A}\Phi|^2-\mu^2|\Phi|^2+\cdots\,,
\fe
where ${\cal Q}$ is the integer charge of $\Phi$.

From this, we learn that the background gauge field $A$ couples to the current
\ie
J_A=-{1\over 2\pi}dc+{\cal Q}\mathscr{J}(\Phi)\,,
\fe
where in $\mathscr{J}(\Phi)$ we use the covariant derivative with the gauge field $c+{\cal Q}A$.  (If we added to the Lagrangian another $AdA$ counterterm, $J_A$ would have had an additional $dA$ term.)  This operator can be simplified using the equations of motion of ${\frak a}$ and $c$
\ie
&\left(2N-{k\over 2}\right)d(\Tr {\frak a})+  dc=0\\
&d(\Tr {\frak a})=-dA+2\pi \mathscr{J}(\Phi)\,,
\fe
to find
\ie
&J_A= {n_v\over 2} \mathscr{J}(\Phi)+\left({k\over 2}-2N\right){dA\over 2\pi}\\
&n_v= -k+4N+2{\cal Q}\,.
\fe
Compare with \eqref{JAJbo} and recall that $n=2$.  As a check, $n_v=k\bmod 2$ is consistent with \eqref{QAaSU2}.

For even $k$, there is the option to have ${\cal Q}={k\over 2}-2N $ such that on-shell, $J_A$ is a trivial operator.

However, for odd $k$, this cannot be done.  Then, we have our regular options.  We can extend $\U_A$, i.e., we write $A=2\hat A $ with a standard $\U$ gauge field $\hat A$, such that ${\cal Q}$ can be half-integer, and the current can vanish on-shell.  Alternatively, $J_A$ does not vanish on-shell and the operator $\cal O$ carries non-zero $\U_A$ charge.

\section{Conclusions}\label{conclusions}

\subsection{Summary}

We started in Section \ref{twoTQFTs} with a TQFT $\cal T$ with a one-form symmetry $\bZ_n^{(1)}$ generated by an anyon $a$ with anomaly $p\bmod 2n$.  Viewing $p$ as a  nonzero arbitrary integer, we discussed a map to another TQFT $\cal T'$
\ie\label{TTpmapp}
{\cal T}\xrightarrow{\ a,\ p\ }{\cal T'}={{\cal T}\boxtimes \U_{-pn}\over \bZ_n} = {{\cal T}\boxtimes \U_{-pn}\over (a,p)}\,,
\fe
where the labels $a$ and $ p$ over the arrow denote that the map depends on them (and not only on $p\bmod 2n$).

We discussed three interpretations of the map \eqref{TTpmapp}.  (See also \cite{Cheng:2022nds}.)
\begin{itemize}
    \item The obvious interpretation is that \eqref{TTpmapp} is obtained by tensoring another TQFT, $\U_{-pn}$, such that we can gauge the anomaly-free diagonal $\bZ_n^{(1)}$.
    \item In Section \ref{Cppgau}, we enriched $\cal T$ by a background $\U$ gauge field, using $a$ as the vison, added an SPT coupling for that $\U$ gauge field, and then turned it into a dynamical field, which we denote by $\frak b$.  This led to $\cal T'$.  
    \item In Section \ref{heirarchycon}, we described how $\cal T'$ can also be thought of as created out of $\cal T$ using the hierarchy construction \cite{HaldaneHierarchy, HalperinHierarchy}.
\end{itemize}
Given the map \eqref{TTpmapp}, we discussed the inverse map by coupling the system to another $\U$ gauge field, which we denoted by $c$ with the coupling ${1\over 2\pi}{\frak b}dc$. This led us in Section \ref{interpretationI} to the duality relation \eqref{Cpp}.

These operations of coupling to a background field, adding an SPT, making it dynamical, and then going back to the original theory, are essentially identical to the $S$ and $T$ operations in \cite{Witten:2003ya}.  These are maps between distinct theories, rather than being phase transitions within the same theory.

One motivation for this exploration was the study of a topological order, which in the far IR is described by the TQFT $\cal T$.  Then, we assume that an Abelian anyon $a$ becomes light at a phase boundary.  And we would like to determine its coupling to $\cal T$ and the phase transition it can trigger.  See Figure \ref{threescales}.

Therefore, in Section \ref{nontoptrant}, we extended our view of the problem by constructing an explicit non-topological quantum field theory that implements the abstract map \eqref{TTpmapp} and its inverse by a phase transition.  That field theory has two gapped phases described by $\cal T$ and $\cal T'$.

Assuming that we add to the TQFT only a single complex (or real) field $\Phi$ that creates $a$, our construction is essentially unique.  To see that, we note that $\Phi$ can couple only to an Abelian gauge field $c$, which we take to be $\U$.  This means that its covariant derivative is $D_c\Phi$.  Then, in order to couple $c$ to the TQFT fields, which we denote generically as $\beta$, we introduce a Lagrange multiplier $\U$ gauge field $\gamma$, and add the coupling
\ie\label{generalcoupl}
{1\over 2\pi} \big(c -C(\beta)\big)d\gamma\,,
\fe
where $C(\beta)$ is a function of the TQFT fields. We choose it such that $\exp\left(i\int C(\beta)\right)$ represents the worldline of the anyon $a$, which generates $\bZ_n^{(1)}$.

One way to think about the theory with the coupling \eqref{generalcoupl} is as follows.  Ignoring $\gamma$, we have two decoupled sectors: $\cal T$ that depends on $\beta$, and the $(\Phi, c)$ system.   Next, we consider $\gamma$ as a classical background field. We recognize it as a field that enriches $\cal T$ through its $\bZ_n^{(1)}$ generated by the anyon $\exp\left(i\int C(\beta)\right)$, and it couples to the $(\Phi, c)$ system as a background field for its magnetic $\U$ global symmetry.  Finally, we make $\gamma$ dynamical and couple the two sectors.

This reasoning shows that our construction is essentially unique.

When $\Phi$ is massive, the low-energy theory is determined by integrating out $\Phi$, $\gamma$, and $c$.  We end up with $\cal T$, and the massive field $\Phi$ creates the TQFT anyon $a$.

When $\Phi$ acquires an expectation value, it Higgses $c$, which means that the coupling \eqref{generalcoupl} sets $C(\beta)$ to be trivial, leading to the TQFT $\cal T'$.

The case with $p=0$, which was discussed in Section \ref{piszero}, is particularly interesting.  Here, $\cal T'$ is not a TQFT.  And, related to that, our transition field theory leads to a gapless phase after the transition.  In this case, there is a natural deformation of the theory that involves adding a local operator $\cal O$ that gaps that phase.  This replaces $\cal T'$ by ${{\cal T}\over \bZ_n}={{\cal T}\over (a)}$, i.e., the transition leads to a phase where the anomaly-free $\bZ_n^{(1)}$ symmetry of $\cal T$ is gauged. In this case, our transition field theory coincides with the unpublished construction of \cite{receipe}.

The construction of \cite{receipe} starts with $\cal T$, then gauges its $\bZ_n^{(1)}$, thus leading to a $\bZ_n^{(0)}$ dual symmetry.  Gauging that $\bZ_n^{(0)}$ symmetry brings us back to $\cal T$, with a dynamical $\bZ_n^{(0)}$ gauge field, which we denoted as ${\mathbf c}^{(1)}$.  Then, the transition theory is the Higgs theory of ${\mathbf c}^{(1)}$.  Our procedure is a generalization of this construction, which works for all values of $p$.  Instead of starting with gauging $\bZ_n^{(1)}$, which cannot be done for $p\ne 0 \bmod 2n$, we enrich $\cal T$ by $\U_X^{(0)}$ using its $\bZ_n^{(1)}$ symmetry, and add an SPT term ${N\over 2\pi}XdX$.  Then, we make $X$ dynamical.  At this stage, the procedure of \cite{receipe} led to a dual $\bZ_n^{(0)}$ symmetry.  Instead, we have a dual $\U^{(0)}$ symmetry.  We gauge it using the gauge field $c$.  Then, we proceed as in \cite{receipe}, to find the transition theory as the Higgs theory of $c$. 

In Section \ref{dual theory}, we used particle/vortex duality to find another description of our theory. For nonzero $p$, it leads to the inverse of the map  \eqref{TTpmapp}
\ie\label{invserTprime}
{\cal T}\xleftarrow{\ \tilde a,\ \tilde p\ }{\cal T'}\,.
\fe
At the level of the TQFT, this inverse map was discussed in Section \ref{inversemap}.
This means that when both $p$ and $\tilde p$ are nonzero, the two TQFTs can be viewed as hierarchy constructions of each other.

In Section \ref{enrichmentsec}, we enriched $\cal T$ by a $\U_A$ global symmetry.  Importantly, we showed that when the $\U_A$ charge of $a$ 
\ie
Q_A(a)={n_v\over n}\bmod 1
\fe
is not an integer, i.e., when $n_v\ne 0 \bmod n$, $\U_A$ cannot act as pure enrichment in our transition field theory.  One implication of that is that for $p=0$, where we had a transition description of gauging $\bZ_n^{(1)}$, that transition is incompatible with the $\U_A$.

More generally, even though the coupling of the background field $A$ of $\U_A$ to $\cal T$ depends only on $n_v\bmod n$, the coupling of $A$  to the transition theory depends on the integer $n_v$.  In particular, it sets the charge of the operator $\cal O$ 
\ie
Q_A({\cal O})=n_v\,.
\fe
Then, even when $n_v=0\bmod n$, nonzero $n_v$ means that $\U_A$ is not a pure enrichment.  Then, for $p=0$, we cannot deform the system, and the negative $\mu^2$ theory is gapless with spontaneously broken $\U_A$. 

To summarize, in addition to the data in $\cal T$ and the anyon $a$, the transition theory depends on two integers $p$ and $n_v$.  In the TQFT, only  $p\bmod 2n$ and $n_v\bmod n$ are relevant.  They determine the spin and the $\U_A$ charge of $a$ modulo one
\ie
&h(a)={p\over 2n}\bmod 1\\
&Q_A(a)={n_v\over n}\bmod 1\,.
\fe
In the transition theory, both $p$ and $n_v$ are relevant as integers.

\subsection{Outlook}

Our work points to several extensions.

We discussed bosonic theories.  The extension to fermionic theories is straightforward.  Of particular interest is the extension to what can be called electronic theories.  These are fermionic theories coupled to $\U_A$, such that they can be placed on spin$^c$ manifolds.  See \cite{metlitski2015s,seiberg2016gapped,seiberg2016duality,Cheng:2025ube} for a description for physicists. In these papers, the spin/charge relation of systems of electrons was expressed as the spin$^c$ condition.

Once fermionic theories are understood, one can couple them to fermionic degrees of freedom.  Again, such theories can be important in the study of electrons.  Here, we can relate different models using various boson/fermion dualities\cite{Barkeshli:2012rja,Karch:2016sxi,seiberg2016duality}),

We emphasized that $\U_{\cal B}$ can have a nontrivial subgroup $\bZ_\ell^{\cal B}$ that does not act on point operators. This subgroup acts on lines.  Or more generally, a subgroup of it, known as a soft symmetry, does not act on line operators, but only on junctions of lines \cite{Davydov2014Bogomolov, Kobayashi:2025ykb}.  It would be nice to explore such examples in more detail. 

We limited ourselves to $\Phi$ in a one-dimensional representation of the gauge group.  Additional transitions can be found by studying $\Phi$ in a higher-dimensional multiplet.   Here, we expect to find also transitions corresponding to gauging non-invertible one-form symmetries \cite{Kong:2013aya, Kaidi:2021gbs,Yu:2021zmu,Roumpedakis:2022aik, Cordova:2023jip}. 

By varying the parameters in the potential as a function of a spatial coordinate $x$, we can create an interface such that for $x<0$, the theory is in $\cal T$, and for $x>0$, it is in $\cal T'$.  It is clear that the interface theory is a $c=1$ chiral CFT.  It would be nice to explore it in more detail to better understand the dynamics of our transition theory.

\section*{Acknowledgments}
We benefited from discussions with Maissam Barkeshli, Xie Chen, Clay Cordova, Dan Freed, Po-Shen Hsin, Wenjie Ji, Ho Tat Lam, John McGreevy, Gregory Moore, Seth Musser, Abhinav Prem, Marvin Qi, Brandon Rayhaun, Amir Raz, Shu-Heng Shao, Sahand Seifnashri, Ashvin Vishwanath, and Edward Witten.  We are particularly thankful to Senthil Todadri for many discussions and for his participation in the early stages of this project. MC is partially supported by NSF grant DMR-2424315.  NS is partially supported by DOE grant DE-SC0009988 and the Simons Collaboration on Ultra-Quantum Matter, which is a grant from the Simons Foundation (Grant No. 651444).

\appendix

\section{Gauging a one-form symmetry in $2+1$d TQFTs}\label{sec:gauging-one-form}

In this Appendix, we will review the process of gauging a one-form global symmetry in a $2+1$d system. In particular, we will apply this procedure to $2+1$d TQFTs.   This gauging can be described from various complementary perspectives.  Therefore, we would like to begin by reviewing some historical background.

\subsection{Historical background}

This procedure was first found in the study of two-dimensional rational conformal field theory.  Specifically, the search for different modular invariants of theories based on the same chiral algebra has led to theories with extended chiral algebras \cite{Cardy:1986ie,Gepner:1986wi,Cappelli:1986hf,Gepner:1986ip,Cappelli:1987xt,Bernard:1986xy,Bouwknegt:1986gc,Altschuler:1987gm,Moore:1988ss,Schellekens:1989am}.  

Another perspective of this construction arose from the Chern-Simons description of $2+1$d TQFT \cite{Witten:1988hf}.  Here, it appears when the gauge group of the theory is replaced by a quotient of it \cite{Moore:1989yh}.  For example, it maps $\U_{2mn^2}\to \U_{2m}$ and ${\rm SU}(2)_{4k}\to {\rm SO}(3)_{4k}$. Importantly, certain non-Abelian chiral algebra extensions, such as ${\rm SU}(2)_{10}\to {\rm Spin}(5)_1$ and ${\rm SU}(2)_{28}\to ({\rm G}_2)_1$,  cannot be understood this way.

However, in all these cases, the more abstract description of these theories in terms of modular tensor categories (MTC) \cite{Moore:1988qv,Moore:1989vd, BakalovKirillov2001, Turaev2010} allows a description of this procedure using the same three-step process  \cite{Moore:1989yh, kirillov2002q, Kong:2013aya, Eliens:2013epa, neupert2016boson,  Yu:2021zmu}, which we will review below.

From a more modern perspective, this process can be viewed as gauging a one-form global symmetry \cite{Gaiotto:2014kfa}.  Then, the simpler cases of taking the quotient of the gauge group correspond to gauging an invertible one-form symmetry. And the more complicated cases like ${\rm SU}(2)_{10}\to {\rm Spin}(5)_1$ and ${\rm SU}(2)_{28}\to ({\rm G}_2)_1$ correspond to gauging a non-invertible one-form symmetry \cite{Kong:2013aya, Kaidi:2021gbs,Yu:2021zmu,Roumpedakis:2022aik, Cordova:2023jip}.

 Gauging amounts to summing over gauge fields.  For a discrete, invertible, one-form symmetry in $2+1$d, the gauge fields are 2-form gauge fields, or more precisely, 2-cocycles, and they can be described as a network of line defects.  This means that the gauging corresponds to summing over all possible insertions of these line defects \cite{Gaiotto:2014kfa,Hsin:2018vcg}, thus relating to the picture in \cite{Moore:1989yh} as proliferation of these lines.   Starting with \cite{Kirillov:2001ti, Frohlich:2003hm,Kong:2013aya}, this picture of proliferation was also generalized to the gauging of non-invertible symmetries. In this note, we limit ourselves to invertible symmetries, leaving the extension to non-invertible symmetries to future work.

In \cite{Bais:2008ni}, this proliferation of lines was described as ``condensation.'' (In \cite{Moore:1989yh,Moore:1991ks}, a closely related phenomenon had been described as condensation of lines.)   Since then, it has become common to refer to this gauging as condensation.  It should be stressed, however, that this process is in no sense a condensation.  First, it is a map between two theories, rather than a transition between them.  Second, no global symmetry is being spontaneously broken here.  And third, no particle condenses (or even proliferates); only their worldlines proliferate.  (See footnote \ref{nocondensation}.)

While we have focused on TQFTs, the same operation can be performed in any theory with anomaly-free one-form symmetry.

\subsection{The procedure}
 \label{sec:three-step}

Here, we review the gauging procedure in the context of an abstract TQFT (or MTC).

A subset of the anyons in the TQFT $\cal T$ is the Abelian anyons ${\cal A}\subset {\cal T}$. They are the symmetry lines of the invertible one-form global symmetry of $\cal T$.  We consider a subgroup of them ${\cal A}_0\subset {\cal A}$.\footnote{We denote by ${\cal A}_0$ both the set of anyons and the Abelian group.} The subgroup must be anomaly-free, meaning that all the anyons in ${\cal A}_0$ are bosonic \cite{Moore:1989yh,Gaiotto:2014kfa,Hsin:2018vcg}.  For simplicity, we assume that ${\cal A}_0$ is a single cyclic group, $\bZ_n$.  Otherwise, the third step below can be more complicated.

Gauging the symmetry is done by proliferating the symmetry defects in spacetime, i.e., summing over all configurations of symmetry defect insertions. Here, the defects in spacetime are the line operators of ${\cal A}_0$.   Therefore, the gauged theory is obtained by summing over all possible insertions of ${\cal A}_0$ lines in $\cal T$.  We denote the gauged TQFT as $\cT\over \cA_0$.

For simplicity, let us assume that $\cA_0$ is a cyclic group and gauge it.  (If $\cA_0$ is a product of cyclic groups, we perform the procedure below sequentially for each factor.) Then, the spectrum of anyons in $\cT\over \cA_0$ can be obtained through the following three-step procedure \cite{Moore:1989yh}:
\begin{enumerate}
\item  First, all the lines charged under ${\cal A}_0$ are projected out. Only the lines with trivial braiding with ${\cal A}_0$ survive. This step is the same as imposing Gauss's law in a gauge theory.  This is expected from the general definition of gauging, which projects onto gauge-invariant operators. We can also see it in the picture of the proliferation of defect lines.  Consider inserting a line $x\in {\cal T}$ with nontrivial braiding with ${\cal A}_0$.  Then, by summing over all possible ${\cal A}_0$ loops linked with the $x$ line, the partition function vanishes.    The elimination of line operators that are not invariant under ${\cal A}_0$ can also be interpreted as due to an anomaly.  Before the gauging, these lines had an 't Hooft anomaly with ${\cal A}_0$.  After the gauging, this anomaly became an ABJ anomaly that explicitly breaks this symmetry.
\item Next, we arrange the lines braiding trivially with ${\cal A}_0$ into orbits under fusing with $\cA_0$. Since all possible insertions of ${\cal A}_0$ are summed over, lines in the same orbit must be identified.  This identification is also standard in gauge theories.  Since the symmetry lines $a\in {\cal A}_0$ are trivial in the gauge theory, the line $x\in {\cal T}$ and the line $a\times x\in {\cal T}$ are indistinguishable in ${\cal T}\over {\cal A}_0$.
\item There is an additional subtlety when a line $x$ is invariant when fused with nontrivial elements of ${\cal A}_0$. Denote the stabilizer subgroup $\cA_x$ as $\{a\in {\cal A}_0\,|\, a\times x=x\}$. Then in ${\cal T}\over {\cal A}_0$, there are multiple descendants of the line $x$, which are associated with lines $a\in\cA_x$ ending on the lines of the anyon $x$.  All of these lines have the same topological spin as $x$. The existence of these states is also familiar in other examples of gauging.  In the context of orbifolds (which is a particular case of gauging), these new lines correspond to twisted states.  And this is exactly how they arise in the Chern-Simons description of this gauging \cite{Moore:1989yh}.
The computation of other observables, such as the S matrix, is more involved. More details can be found in, e.g., \cite{Eliens:2013epa}.
\end{enumerate}

It is worth emphasizing that the prescription does not rely on the theory $\cal T$ being topological. The same procedure can be applied to a non-topological theory with a non-anomalous one-form symmetry.

\subsection{Dual symmetry}
\label{app:dualsymmetry}

Gauging a global symmetry introduces a dual (quantum) symmetry. When gauging a discrete one-form symmetry $\cA_0^{(1)}$ in $2+1$d, the new theory has a dual $\tilde{\cA}_0^{(0)}$ 0-form symmetry,\footnote{Mathematically, $\tilde{\cA}_0^{(0)}$ is the Pontryagin dual of ${\cA}_0^{(1)}$, i.e., $\tilde{\cA}_0^{(0)}={\rm Hom}(\cA_0^{(1)},\U)$.} whose symmetry operators are given by the Wilson surfaces (holonomies) of the gauge field of $\cA_0^{(1)}$ over two-cycles. In some contexts (e.g., in the study of orbifolds), this dual symmetry is called a ``quantum symmetry.''

It is well-known that the procedure can be inverted: starting with ${\cal T}\over \cA_0$, and then gauging the dual zero-form symmetry, i.e., summing over gauge fields of $\tilde{\cal A}_0^{(0)}$, leads to the original theory.\footnote{ More precisely, there is an ambiguity in the gauging of this dual zero-form symmetry.  In this case, it is associated with adding  Dijkgraaf-Witten terms \cite{Dijkgraaf:1989pz}.}

This dual symmetry is particularly interesting when step 3 in the gauging procedure above is activated.  Then, some elements of  $\tilde{\cA}_0^{(0)}$ permute lines in $\cT\over \cA_0$.  These elements act faithfully on the TQFT.  The elements of $\tilde{\cA}_0^{(0)}$ that do not act faithfully can still be used to enrich $\cT\over \cA_0$   \cite{etingof2010fusion,barkeshli_symmetry_2019, Fidkowski:2016svr}. 

\subsection{Enrichment}\label{enrichementapp}

We now consider what happens when the theory before gauging is coupled to a background gauge field for a zero-form symmetry $G^{(0)}$. For simplicity, we assume that none of the elements in $G^{(0)}$ act faithfully in the TQFT, i.e., the $G^{(0)}$ gauge fields couple to the TQFT only via enrichment. (The more general case and additional details are discussed in  \cite{Bischoff:2018juy}.)  The enrichment can be described by assigning $G^{(0)}$ projective quantum numbers to the anyons. 

If the anyons in $\cA_0^{(1)}$ carry nontrivial projective  $G^{(0)}$ quantum numbers, then after gauging $\cA_0^{(1)}$, the resulting TQFT cannot be coupled to background $G^{(0)}$ gauge fields.  Instead, $G^{(0)}$ should be extended by $\tilde{\cA}_0^{(0)}$ to $\tilde G^{(0)}$. We should stress, however, that the gauging still yields a consistent TQFT.

Let us describe this group extension more explicitly. First, before the gauging, the enrichment (or equivalently, the symmetry fractionalization class) is specified by a cohomology class $[\omega]$ in $H^2(G^{(0)}, {\cal A})$   \cite{etingof2010fusion,barkeshli_symmetry_2019, Fidkowski:2016svr}.  One way to phrase it is that the background gauge fields of $G^{(0)}$ lead to a background gauge field $\omega$ for the one-form symmetry $\cal A$.   And the latter couples to the TQFT. (In \cite{Seiberg:2025bqy}, this process was described as a transmutation of the zero-form symmetry $G^{(0)}$ into a one-form symmetry.)

As we said, after gauging $\cA_0^{(1)}$, consistency may force us to extend $G^{(0)}$ to $\tilde{G}^{(0)}$.  The latter is a central extension of $G^{(0)}$ by $\tilde{\cA}_0^{(0)}$,  parametrized by a 2-cocycle $\tilde{\omega}$ in $H^2(G^{(0)}, \tilde{\cA}_0^{(0)})$. This 2-cocycle is determined as follows.  For $g,h\in G^{(0)}$, choose a representative 2-cocycle $\omega(g,h)\in \cA$.  Then $\tilde{\omega}(g,h)\in \tilde{\cA}_0^{(0)}$ is defined using the braiding
\begin{equation}
\label{dualomega}
    [\tilde{\omega}(g,h)](x)=e^{2\pi i \langle x,\omega(g,h)\rangle}\quad , \quad \forall x\in \cA_0^{(1)}.
\end{equation}
It is easy to show that $\tilde{\omega}$ is a 2-cocycle on $G^{(0)}$ valued in $\tilde{\cA}_0^{(0)}$, so defines an element of $H^2(G^{(0)}, \tilde{\cA}_0^{(0)})$. $\tilde{G}^{(0)}$ is then determined by this 2-cocycle.

Let us phrase it in terms of the background fields.  We started with a background $G^{(0)}$ gauge field.  We used it to write an $\cal A$ background gauge field $\omega$. (Since we assumed that $G^{(0)}$ does not act faithfully on $\cal T$, this is the only coupling between the $G^{(0)}$ background gauge field and $\cal T$.)  Then, we gauged ${\cal A}_0^{(1)}$.  One consequence of that is that elements in $\cal T$ with nontrivial braiding with $a\in {\cal A}_0^{(1)} $ are removed.  This can be inconsistent with the original coupling of the $G^{(0)}$ gauge fields via $\omega$, thus forcing us to restrict the allowed $\omega$, which in turn restricts the $G^{(0)}$ gauge fields.  This can be stated as an extension of $G^{(0)}$.

We now specialize to $G^{(0)}=\U_A$, with charge $Q_A$. Suppose the enrichment is through a vison $v$, and ${\cal A}_0^{(1)}$ is generated by an order-$n$ boson $a$. The $\U_A$ charge of $a$ is determined by the braiding of $v$ and $a$ as $Q_A(a)=\langle v,a\rangle=\frac{n_v}{n}\bmod 1$. To find the extension, we first choose a generator $\tilde{a}$ of $\tilde{\cA}_0^{(0)}={\rm Hom}(\cA_0^{(1)},\U)$, defined by the map $\tilde{a}(a)=e^{2\pi i\over n}$.  Then, using \eqref{dualomega}, it is easy to see that the extension is given by $e^{2\pi i Q_A}=\tilde{a}^{n_v}$. The extension is nontrivial as long as $n_v\neq 0\bmod n$. In this case, we can trivialize the extension by defining $Q_{\hat{A}}=\frac{n}{\gcd(n,n_v)}Q_A$, so that $e^{2\pi i Q_{\hat{A}}}=\tilde{a}^{\frac{n_vn}{\gcd(n_v,n)}}=1$. This is equivalent to using $\hat{v}=v^{n\over \gcd(n,n_v)}$ as the vison for the extended symmetry.

\bibliographystyle{JHEP}
\bibliography{Hall.bib}

   \end{document}